\documentclass[12pt]{article}
\usepackage{amssymb}
\usepackage{subfig}
\usepackage{color}
\usepackage{epsfig,amssymb,amsfonts,amsmath,graphicx,dsfont,cite,xfrac}
\usepackage{authblk}
\usepackage{mathabx}
\usepackage{mathtools}

\usepackage{tikz,pgfplots}
\usetikzlibrary{positioning}
\usetikzlibrary{matrix}

\definecolor{mygray}{gray}{0.5}

\usepackage{cite}
\usepackage[colorlinks=true,linkcolor=blue,citecolor=red]{hyperref}

\newcommand{\be}{\begin{equation}}
\newcommand{\ee}{\end{equation}}
\newcommand{\bea}{\begin{eqnarray}}
\newcommand{\eea}{\end{eqnarray}}

\title{Time-dependent rational extensions of the parametric oscillator: quantum invariants and the factorization method}

\author[${1}$]{K. Zelaya\thanks{zelayame@crm.umontreal.ca}}
\author[${1,2}$]{V. Hussin\thanks{hussin@dms.umontreal.ca}}

\affil[${1}$]{\footnotesize Centre de Recherches Math\'ematiques, Universit\'e de Montr\'eal, Montr\'eal H3C 3J7, QC, Canada}

\affil[${2}$]{\footnotesize D\'epartement de Math\'ematiques et de Statistique, Universit\'e de Montr\'eal, Montr\'eal H3C 3J7, QC, Canada}

\date{}

\begin{document}

\maketitle

\begin{abstract}
New families of time-dependent potentials related to the parametric oscillator are introduced. This is achieved by introducing some general time-dependent operators that factorize the appropriate constant of motion (quantum invariant) of the parametric oscillator, leading to new families of quantum invariants that are almost-isospectral to the initial one. Then, the respective time-dependent Hamiltonians are constructed, and the solutions of the Schr\"odinger equation are determined from the intertwining relationships and by finding the appropriate time-dependent complex-phases of the Lewis-Riesenfeld approach. To illustrate the results, the set of parameters of the new potentials are fixed such that a family of time-dependent rational extensions of the parametric oscillator is obtained. Moreover, the rational extensions of the harmonic oscillator are recovered in the appropriate limit.
\end{abstract}

\section{Introduction}
The dynamics of non-relativistic quantum systems is determined from the solutions of the Schr\"odinger equation, where the Hamiltonian operator characterize the system. For a wide variety of systems, it is sufficient to consider stationary models (time-independent Hamiltonians), such as the harmonic oscillator, Hydrogen atom, diatomic and polyatomic interactions, just to mention some examples. In such case, the time-evolution operator allows to decompose the Schr\"odinger into a stationary eigenvalue equation. Even for stationary systems, only some few models are known to admit exact solutions and thus the search of new exactly solvable model becomes a challenging task. Remarkably, the factorization method~\cite{Mie04,Kha93,Car00,Coo01,Don07} becomes an outstanding technique to explore the existence of new solvable stationary models. The latter is achieved by inspecting the eigenvalue equation associated with the Hamiltonian of an already known solvable model and relating it with the Darboux transformation~\cite{Mat91}. In this form, a wide class of new exactly solvable models have been reported in the literature for Hermitian Hamiltonians~\cite{Mie84,Mie00,Fer99}, non-Hermitian Hamiltonians with real spectrum in the $\mathcal{PT}$ and non-$\mathcal{PT}$ regime~\cite{Zno00,Bag01,Ros15,Ros18,Bla18,Cor15} and position-dependent mass models~\cite{Que06,Cru09,Cru13}, just to mention some applications. 

The class of exactly solvable time-dependent Hamiltonians is less known, since in general it is not possible to identify an eigenvalue equation for the Hamiltonian and the existence of an orthogonal set of solutions can not be taken for granted. To address these systems, sometimes we have to rely on approximation techniques such as the sudden and the adiabatic approximations~\cite{Sch02}. Therefore, the search of exactly solvable time-dependent models becomes a relevant task in the field of mathematical physics. Despite the complexity, time-dependent phenomena find interesting applications in electromagnetic traps of charged particles~\cite{Pau90,Com86,Pri83,Gla92}, and also in classical optics analogs under the paraxial approximation~\cite{Cru17}, where electromagnetic waves propagate in a spatial-varying dielectric media~\cite{Raz19,Con19}. The parametric oscillator (\textit{nonstationary oscillator}~\cite{Dod95,Zel19}) is perhaps the most well known time-dependent model in quantum mechanics that admits a set of exact solutions. It is characterized by an oscillator-like interaction with a frequency term that depends on time, it was studied by Lewis and Riesenfeld~\cite{Lew68,Lew69} for both classical and quantum systems. Indeed, for the quantum model, a set of orthogonal solutions is found by constructing the appropriate constant of motion (quantum invariant) of the system and solving the spectral problem associated with it. Such an approach has been essential to find the solutions of other time-dependent systems such as the nonstationary oscillator with a singular barrier~\cite{Dod98} and the Caldirola-Kanai oscillator~\cite{Gue15}. 

In this text, we explore the construction of new exactly solvable models associated with the parametric oscillator. To this end, we take advantage of the nonstationary spectral problem associated with one of the constants of motion of the system. The factorization method is then implemented to generated new families of operators that share their spectral properties with the initial invariant. These new operators are not quantum invariants of the parametric oscillator, but it can be shown that they are constants of motion of some new families of time-dependent Hamiltonians. The solutions of the parametric oscillator are thus mapped into solutions of the Schr\"odinger equation associated with the new Hamiltonians. 

In previous works, an alternative approach to construct time-dependent models was proposed by Bagrov and Samsonov~\cite{Bag96}, where the intertwining relations between two Schr\"odinger equations are introduced in order to generate new exactly solvable models. The method by itself does not provide information about the constants of motion of the system. Moreover, it is not granted that the mapping of an orthogonal set of solutions from the initial model would lead to an orthogonal set of solutions for the new model, see~\cite{Zel17}. Nevertheless, some interesting new exactly solvable models has been constructed in this way~\cite{Zel17,Con17,Cru19,Cen19}. From our approach, the constants of motion are obtained by construction and the orthogonality of the new set of solutions is inherited from the initial parametric oscillator model and the factorization operators.

The organization of this paper is as follows. In Sec.~\ref{sec:PO} we briefly summarize the approach of Lewis-Reisenfeld~\cite{Lew69}. In Sec.~\ref{sec:FAC}, we implement the factorization method to the spectral problem associated with the appropriate quantum invariant of the parametric oscillator. The latter allows to construct quantum invariants that are associated with some new families of time-dependent Hamiltonians. The mechanism to generate the respective solutions is also discussed. In particular, the rational extensions of the parametric oscillator for both the one-step and two-step factorizations are presented in Sec.~\ref{sec:REX}, the solutions of which are expressed in terms of a family of time-dependent exceptional Hermite polynomials. To illustrate our results, in Sec.~\ref{sec:EXA} we consider two specific frequency profiles, chosen in such a way that both periodic and non-periodic solutions are obtained. In addition, we show that the stationary rational extensions of the harmonic oscillator, reported in~\cite{Mar13,Gom14}, are recovered in the appropriate limit. App.~\ref{sec:APPC} contains the explicit calculation of the time-dependent complex-phase required in the Lewis-Riesenfeld approach to determine the solutions of the Schr\"odinger equation. In App.~\ref{sec:APPB} we show the explicit calculations used to determine the time-dependent Hamiltonians from their respective constants of motion.

\section{Time-dependent parametric oscillator}
\label{sec:PO}
The Hamiltonian associated with the parametric oscillator is defined through the quadratic form
\begin{equation}
H_{0}(t)=\hat{p}^{2}+\Omega^{2}(t)\hat{x}^{2}+F(t)\hat{x}+\mathcal{V}(t)\hat{\mathbb{I}} \, , \quad [\hat{x},\hat{p}]=i\hbar \hat{\mathbb{I}} \, ,
\label{eq:PO1}
\end{equation}
where $\hat{x}$ and $\hat{p}$ stand for the canonical position and momentum observables, respectively. The term $F(t)$ represents a driving force and $\mathcal{V}(t)$ a time-dependent zero-point energy. To simplify the notation, the identity operator $\hat{\mathbb{I}}$ will be omitted each time it multiplies a time-dependent function or a constant. Without loss of generality, we consider a null zero-point energy term (see Sec.~\ref{subsubsec:SHIN}). The solutions of the parametric oscillator~\eqref{eq:PO1} are determined from the Schr\"odinger equation
\begin{equation}
i\frac{\partial}{\partial t}\psi^{(0)}(x,t)=H_{0}(t)\psi^{(0)}(x,t)\, .
\label{eq:PO2}
\end{equation}
Given that the Hamiltonian depends explicitly on time, Eq.~\eqref{eq:PO2} can not be decomposed into an eigenvalue equation for $\hat{H}_{0}(t)$. Nevertheless, a set of orthogonal solutions can still be found following the approach proposed by Lewis and Riesenfeld~\cite{Lew68,Lew69}, where the appropriate constant of motion (\textit{invariant operator or quantum invariant}) of the system, computed from the condition
\begin{equation}
\frac{d}{dt}I_{0}(t)=i[H_{0}(t),I_{0}(t)]+\frac{\partial}{\partial t}I_{0}(t) = 0 \, ,
\label{eq:INVI0}
\end{equation}
admits a nonstationary eigenvalue equation of the form
\begin{equation}
I_{0}(t)\varphi^{(0)}_{n}(x,t)=\lambda^{(0)}_{n}\varphi^{(0)}_{n}(x,t) \, ,
\label{eq:PO5}
\end{equation}
with $\varphi_{n}(x,t)$ the $n$-th nonstationary eigenfunction and $\lambda_{n}^{(0)}$ the respective time-independent eigenvalue~\cite{Lew69}. For $F(t)=0$, the invariant operator has been found with the aid of an appropriate ansatz~\cite{Lew69}, whereas for $F(t)\neq 0$ it has been found through the use of geometrical transformations~\cite{Zel19b,Zel19}. For the case under consideration we have
\begin{multline}
I_{0}(t)=\sigma^2\hat{p}^2+\left( \frac{\dot{\sigma}^{2}}{4}+\frac{1}{\sigma^2} \right)\hat{x}^2-\frac{\sigma\dot{\sigma}}{2}(\hat{x}\hat{p}+\hat{p}\hat{x})+\sigma \mathfrak{W} \hat{p}+\left(2\frac{\gamma}{\sigma^{2}}-\frac{\mathfrak{W}\dot{\sigma}}{2} \right)\hat{x} + \left( \frac{\gamma^2}{\sigma^2} + \frac{\mathfrak{W}^{2}}{4} \right) \, , 
\label{eq:PO3}
\end{multline} 
where $\dot{z}=dz/dt$ stands for the time derivative. The real-valued functions $\sigma(t), \gamma(t)$ and $\mathfrak{W}(t)$ are determined from
\begin{equation}
\ddot{\sigma}+4\Omega^{2}(t)\sigma=\frac{4}{\sigma^{3}} \, , \quad \ddot{\gamma}+4\Omega^{2}(t)\gamma=2 F(t) \, , \quad \mathfrak{W}=\sigma\dot{\gamma}-\dot{\sigma}\gamma \, .
\label{eq:PO4}
\end{equation}
Notice that $\sigma(t)$ solves the nonlinear Ermakov equation~\cite{Erm08,Pin50}, which has been studied extensively in the literature (for a detailed discussion see also~\cite{Ros15,Bla18}). A general solution is provided by the nonlinear superposition
\begin{equation}
\sigma^{2}(t)=a q_{1}^{2}(t)+bq_{1}(t)q_{2}(t)+c q_{2}^{2}(t) \, , \quad b^{2}-4ac=-\frac{16}{w_{0}^{2}} \, , \quad a,c>0 \, ,
\label{eq:PO41}
\end{equation}
where $q_{1,2}$ are two linearly independent solutions of the linear equation
\begin{equation}
\ddot{q}_{1,2}+4\Omega^{2}(t)q_{1,2}=0  \, ,
\label{eq:PO42}
\end{equation}
with the Wronskian $W(q_1,q_2)=w_{0}\neq 0$ in general a complex constant. The constraints on $a,b,c$ given in~\eqref{eq:PO41} ensure that $\sigma^{2}(t)>0$ at each time. From~\eqref{eq:PO42} it follows that $q_{1}$ and $q_{2}$ correspond to two linear independent solutions of the classical equation of motion for the parametric oscillator. In turn, from~\eqref{eq:PO4}, $\gamma(t)$ solves the classical parametric oscillator subjected to an external driving force $2F(t)$.

The eigenfunctions $\varphi_{n}^{(0)}(x,t)$ are computed with the help of the factorization $I_{0}(t)=A^{\dagger}A+1$, where the operators $A$ and $A^{\dagger}$ are introduced, in analogy to the boson ladder operators~\cite{Dir35}, as a combination of linear terms in $\hat{x}$ and $\hat{p}$ with time-dependent coefficients. After some calculations we find 
\begin{equation}
A=\sigma\frac{\partial}{\partial x}+\left( \frac{1}{\sigma}- i \frac{\dot{\sigma}}{2} \right) x + \left(\frac{\gamma}{\sigma}+ i\frac{\mathfrak{W}}{2}\right) \, , \quad  A^{\dagger}=-\sigma\frac{\partial}{\partial x}+\left( \frac{1}{\sigma} + i \frac{\dot{\sigma}}{2} \right) x + \left(\frac{\gamma}{\sigma} - i\frac{\mathfrak{W}}{2}\right) \, ,
\label{eq:PO10}
\end{equation}
where in~\eqref{eq:PO10} we have used the coordinate representation $\hat{x}\rightarrow x$ and $\hat{p}\rightarrow \frac{1}{i}\frac{d}{dx}$. On the other hand, straightforward calculations lead to the following commutation rules:
\begin{equation}
[A,A^{\dagger}]=2 \, , \quad [I_{0},A]=-2A \, , \quad [I_{0},A^{\dagger}]=2A^{\dagger} \, .
\label{eq:POCOM}
\end{equation}
Therefore, $A$ and $A^{\dagger}$ are ladder operators for the eigenfunctions of the invariant operator $I_{0}(t)$,
\begin{equation}
A\varphi^{(0)}_{n+1}=\sqrt{2n+1} \, \varphi^{(0)}_{n} \, , \quad A^{\dagger}\varphi^{(0)}_{n}=\sqrt{2n+1} \, \varphi^{(0)}_{n+1} \, , \quad n=0,1,\cdots \, .
\label{eq:PO9}
\end{equation}
The eigenfunction $\varphi_{0}^{(0)}(x,t)$ is computed, in analogy to the stationary oscillator case, from the condition $A\varphi_{0}^{(0)}=0$. For $n\neq 0$, the eigenfunctions $\varphi_{n}^{(0)}$ are obtained from the iterated action of $A^{\dagger}$ on $\varphi_{0}^{(0)}$. We thus get~\cite{Zel19b} 
\begin{equation}
\varphi^{(0)}_{n}(x,t) = \sqrt{\frac{1}{2^{n}n!\sqrt{\pi}}}  \, \exp\left[ \left(-\frac{1}{2}+i\frac{\dot{\sigma}\sigma}{4} \right)\left(\frac{x+\gamma}{\sigma}\right)^{2} - i \frac{\dot{\gamma}}{2}x \right] 
 \frac{1}{\sqrt{\sigma}} \mathtt{H}_{n}\left(\frac{x+\gamma}{\sigma} \right)  \, ,
\label{eq:PO7}
\end{equation}
with $\mathtt{H}_{n}(z)$ the \textit{Hermite polynomials}~\cite{Olv10} and $\varphi_{n}^{(0)}(x,t)$ associated with the eigenvalue
\begin{equation}
\lambda_{n}^{(0)}=2n+1 \, \quad n=0,1,\cdots \, .
\label{eq:PO71}
\end{equation}
The orthogonality and normalization of the set $\{\varphi^{(0)}_{n}(x,t) \}_{n=0}^{\infty}$ was computed with respect to the physical inner-product
\begin{equation}
\langle\varphi_{m}^{(0)}(t)\vert\varphi^{(0)}_{n}(t)\rangle=\int_{-\infty}^{\infty}dx \, \left[\varphi_{m}^{(0)}(x,t) \right]^{*}\varphi_{n}^{(0)}(x,t)=\delta_{n,m} \, ,
\label{eq:PO72}
\end{equation}
where $z^{*}$ denotes the complex conjugate of $z$. Eq.~\eqref{eq:PO72} holds provided that both eigenfunctions are evaluated at the same time. In general we have $\langle \psi_{m}^{(0)}(t')\vert\psi_{n}(t)\rangle\neq \delta_{n,m}$ for $t\neq t'$. 

It is worth to remark that $\varphi^{(0)}_{n}(x,t)$ is not a solution to the Schr\"odinger equation~\eqref{eq:PO2}. However, we can identify a complete set of solutions after finding the appropriate time-dependent complex-phase~\cite{Lew69} of the form
\begin{equation}
\psi_{n}^{(0)}(x,t)=e^{i\chi_{n}^{(0)}(t)}\varphi^{(0)}_{n}(x,t) \, ,
\label{eq:PO12}
\end{equation}
where $\chi^{(0)}_{n}(t)$ is computed after substituting~\eqref{eq:PO12} in~\eqref{eq:PO2}. It leads to
\begin{equation}
\dot{\chi}_{n}^{(0)}(t)=\langle \varphi_{n}^{(0)}(t)\vert i\frac{\partial}{\partial t}-H_{0}(t)\vert\varphi_{n}^{(0)}(t)\rangle \, .
\label{eq:PO51}
\end{equation}
The complex-phases $\chi_{n}^{(0)}(t)$ have been found in the literature using several methods~\cite{Lew69,Zel19b,Zel19}. For the case under consideration it is given as (see App.~\ref{sec:APPC})
\begin{equation}
\begin{aligned}
\chi^{(0)}_{n}(t)=-\frac{\gamma\dot{\gamma}}{4}+\frac{1}{2}\int^{t}dt'F(t')\gamma(t')-(2n+1)\int^{t}\frac{dt'}{\sigma^{2}(t')} \, , 
\end{aligned}
\label{eq:PO6}
\end{equation}
where the last integral in~\eqref{eq:PO6} can be expressed in terms of $q_{1,2}(t)$ in~\eqref{eq:PO42} as~\cite{Bla18}
\begin{equation}
\int^{t}\frac{dt'}{\sigma^{2}(t')}=\frac{1}{2}\arctan\left[ \frac{w_0}{4}\left( b+2c\frac{q_2}{q_1} \right) \right] \, .
\label{eq:PO6-1}
\end{equation}

An additional property of the ladder operators $A$ and $A^{\dagger}$ is given by their relation with the canonical position and momentum operators $\hat{x}$ and $\hat{p}$, respectively. From~\eqref{eq:PO10} we find
\begin{equation}
\begin{aligned}
& \hat{x}=\frac{\sigma}{2} \left( A+A^{\dagger} \right)-\gamma \, , \quad \hat{p}=\frac{1}{2}\left( \Xi A + \Xi^{*} A^{\dagger} \right) - \frac{\dot{\gamma}}{2} \, , \quad \Xi(t)=-\frac{i}{\sigma}+\frac{\dot{\sigma}}{2} \, .
\end{aligned}
\label{eq:PO11}
\end{equation}
From this, expectation values for the physical quadratures can be computed with ease, as we discuss in the following section.

\subsection{Harmonic oscillator limit}
\label{subsec:GCS}
From the parametric oscillator Hamiltonian~\eqref{eq:PO1} it is clear that a constant frequency and a null driving force lead us to the harmonic oscillator Hamiltonian. That is, for $\Omega^{2}(t)=1$ and $F(t)=0$ we obtain
\begin{equation}
H_{0}(t)\vert_{\substack{\Omega(t)=1 \\ F(t)=0}}\rightarrow \,  H=\hat{p}^{2}+\hat{x}^{2} \, .
\label{eq:HO1}
\end{equation}
In such case, the invariant operator $I_{0}(t)$ is still a time-dependent operator and it becomes a constant of motion of $H$. Moreover, the solutions of $\sigma$ and $\gamma$, computed from~\eqref{eq:PO4}, are given as
\begin{equation}
\sigma^{2}(t)=\frac{a+c}{2}+\frac{a-c}{2}\cos 4t +\sqrt{ac-1}\,\sin 4t  \, , \quad \gamma(t)=\mathcal{A}\cos(2t+\phi) \, ,
\label{eq:HO2}
\end{equation}
with $\mathcal{A}\in\mathbb{R}$ an amplitude of oscillation, $\phi\in[0,2\pi]$ a phase-shift and the constraint $b=2\sqrt{ac-1}$ has been used. Interestingly, for $a,b\neq 1$, the solutions~\eqref{eq:PO12} reduce to
\begin{multline}
\psi_{n}^{(0)}(x,t)=\frac{1}{\sqrt{2^{n}n!\sigma\sqrt{\pi}}}\exp\left\{-\left( -\frac{1}{2}+i\frac{\dot{\sigma}\sigma}{4}\right)\left( \frac{x+\gamma}{\sigma} \right)^{2}-i\frac{\dot{\gamma}}{2}x-i\frac{\dot{\gamma}\gamma}{4}\right\} \\
\times\exp\left\{-\frac{i}{2}\left( 2n+1 \right)\arctan\left(\sqrt{ac-1}+c\tan 2t \right) \right\}\mathtt{H}_{n}\left(\frac{x+\gamma}{\sigma} \right) \, .
\label{eq:HO2-1}
\end{multline}
From~\eqref{eq:HO2-1} it is clear that $n=0$ leads to the wave function of the conventional squeezed states evolving in time, for an explicit expression see~\cite{Ger05,Nie97}. The oscillation amplitude $\mathcal{A}$ and the phase-shift $\phi$ play the role of the modulus and complex-phase of the coherence parameter, respectively. Moreover, the constants $a$ and $c$ give information about the squeezing parameter. On the other hand, for general $n$ and after some calculations, it may be shown that~\eqref{eq:HO2-1} leads to the more general family of \textit{squeezed number states} introduced in~\cite{Nie97}.
 
For the special case $a=c=1$, the solutions~\eqref{eq:HO2} simply become
\begin{equation}
\sigma(t)=1 \, , \quad \gamma(t)=\mathcal{A}\cos(2t+\phi) \, , 
\label{eq:HO3}
\end{equation}
and
\begin{equation}
\mathfrak{W}=\dot{\gamma} \, , \quad \chi_{n}^{(0)}(t)=-\frac{\gamma\dot{\gamma}}{2}-(2n+1)t \, .
\end{equation}
The invariant operator, the eigenfunctions $\varphi_{n}^{(0)}(x,t)$ and the solutions to the Schr\"odinger equation reduce to
\begin{equation}
I_{0}(t)=\hat{p}^2+\hat{x}^2+\dot{\gamma}\hat{p}+2\gamma\hat{x} + \mathcal{A}^{2}  \, ,
\label{eq:HO4}
\end{equation} 
\begin{equation}
\varphi^{(0)}_{n}(x,t)=e^{-\frac{1}{2}\left(x +\gamma\right)^{2}}\sqrt{\frac{1}{2^{n}n!\sqrt{\pi}}}\, \mathtt{H}_{n}\left(x+\gamma\right)  \, , \quad \psi_{n}^{(0)}(x,t)=e^{-i\frac{\gamma\dot{\gamma}}{2}-i(2n+1)t}\varphi_{n}^{(0)}(x,t) \, ,
\label{eq:HO5}
\end{equation}
respectively. After using $\hat{x}$ and $\hat{p}$ given in~\eqref{eq:PO11}, together with the algebraic properties of the ladder operators, we obtain that the mean values $\langle\hat{x}\rangle_{n}=\langle\psi^{(0)}_{n}(t)\vert\hat{x}\vert\psi^{(0)}_{n}(t)\rangle$ and $\langle\hat{p}\rangle_{n}=\langle\psi^{(0)}_{n}(t)\vert\hat{p}\vert\psi^{(0)}_{n}(t)\rangle$ are given by
\begin{equation}
\langle \hat{x} \rangle_{n}=-\gamma=-\mathcal{A}\cos(2t+\phi) \, , \quad \langle \hat{p} \rangle_{n}=-\frac{\dot{\gamma}}{2}=\mathcal{A}\sin(2t+\phi) \, .
\label{eq:HO6}
\end{equation}
From the latter, we can rewrite the solutions~\eqref{eq:HO5} in the convenient form
\begin{equation}
\psi^{(0)}_{n}(x,t)=\exp\left[ -\frac{1}{2}\left(x - \langle \hat{x}\rangle_{n} \right)^{2} + i \langle \hat{p}\rangle_{n}x -i\frac{\langle\hat{x}\rangle_{n}\langle\hat{p}\rangle_{n}}{2} - i\left(2n+1\right)t \right]\frac{\mathtt{H}_{n}\left(x-\langle\hat{x}\rangle_{n}\right)}{\sqrt{2^{n}n!\sqrt{\pi}}} \, . 
\label{eq:HO7}
\end{equation}

For $n=0$, we recover the time evolution of the conventional coherent states of the harmonic oscillator (Glauber coherent states)~\cite{Gla07}. Moreover, for arbitrary $n$ we obtain the more general class of \textit{displaced number states}~\cite{Nie97}, also known as \textit{generalized coherent states}~\cite{Phi14}. 

The variances for the physical position and momentum observables associated with either the displaced~\eqref{eq:HO7} or the squeezed number states~\eqref{eq:HO2-1} can be computed with ease from the relations~\eqref{eq:PO11}, together with the action of the ladder operators. Although, these results will not be discussed in this work, for details see~\cite{Nie97,Phi14}.


\section{New exactly solvable time-dependent potentials}
\label{sec:FAC}
For stationary systems, the factorization method provides an outstanding technique to explore the construction of new exactly solvable models~\cite{Coo01,Mie04,Don07}. The latter is possible since stationary Hamiltonians admit an eigenvalue equation that has the form required for the Darboux transformation~\cite{Mat91}. In turn, for time-dependent models, the method can not be applied in a straightforward way, as mentioned before, no eigenvalue equation is associated with the Hamiltonian. However, by generalizing the set of factorization operators introduced in~\eqref{eq:PO10}, it is possible to factorize
the nonstationary eigenvalue equation~\eqref{eq:PO5}. The inverted order of the factorization leads to new operators whose spectral properties are inherited from $I_{0}(t)$. Moreover, these new operators are the quantum invariants related to some new time-dependent Hamiltonians, where the latter can be constructed with ease. The respective solutions of the Schr\"odinger are determined through the action of the factorization operators on the eigenfunctions of $I_{0}(t)$, and by finding the appropriate complex-phases. 

\subsection{One-step factorization}
\label{subsec:1SF}
In Sec.~\ref{sec:PO} we have introduced a set of ladder operators~\eqref{eq:PO10} that factorize $I_{0}(t)$. In the spirit of the factorization method of the harmonic oscillator~\cite{Mie84}, we introduce a generalized couple of operators of the form 
\begin{equation}
B_{1}:=A+\mathcal{F}_{1}\equiv \sigma\frac{\partial}{\partial x}+\left( \frac{1}{\sigma}-i\frac{\dot{\sigma}}{2} \right)x+\left(\frac{\gamma}{\sigma}+i\frac{\mathfrak{W}}{2} \right) + \mathcal{F}_{1} \, , \quad B_{1}^{\dagger}=A^{\dagger}+\mathcal{F}_{1} \, , 
\label{eq:SPO0}
\end{equation}
with $\mathcal{F}_{1}=\mathcal{F}_{1}(x,t)$ a real-valued function and $\{ A, A^{\dagger} \}$ the set of ladder operators of the parametric oscillator~\eqref{eq:PO10}. The new operators $B_{1}$ and $B_{1}^{\dagger}$ are constructed such that $I_{0}(t)$ factorizes as
\begin{equation}
I_{0}(t)=B^{\dagger}_{1}B_{1}+\epsilon_{1} \, , \quad \epsilon_{1}\leq\lambda^{(0)}_{0} \, .
\label{eq:SPO1}
\end{equation}
After substituting~\eqref{eq:SPO0} into~\eqref{eq:SPO1} and comparing terms with~\eqref{eq:PO3}, we find that $\mathcal{F}_{1}(x,t)$ is solution to the Riccati equation~\cite{Inc56}
\begin{equation}
-\sigma\frac{\partial\mathcal{F}_{1}}{\partial x}+2\left( \frac{x+\gamma}{\sigma} \right)\mathcal{F}_{1}+\mathcal{F}_{1}^{2} +\epsilon_{1}-1=0 \, .
\label{eq:SPO3}
\end{equation}
The latter can be rewritten in a more convenient form by introducing a reparametrization $z(x,t)$ and a function $W_{1}(z)$ of the form
\begin{equation}
z(x,t):=\frac{x+\gamma}{\sigma} \, , \quad \mathcal{F}_{1}(z):=-z+W_{1}(z) \, ,
\label{eq:SPO4}
\end{equation}
such that we recover the well known form of the Riccati equation
\begin{equation}
-\frac{\partial W_{1}}{\partial z}+W_{1}^{2}=z^{2}-\epsilon_{1} \, .
\label{eq:SPO5}
\end{equation}
The latter is linearizable into the eigenvalue equation
\begin{equation}
-\frac{\partial^{2}u_{\epsilon_{1}}}{\partial z^{2}}+z^{2}u_{\epsilon_{1}}=\epsilon_{1}u_{\epsilon_{1}} \, , \quad W_{1}=-\frac{1}{u_{\epsilon_{1}}}\frac{\partial u_{\epsilon_{1}}}{\partial z} \, , \quad u_{\epsilon_{1}}=u_{\epsilon_{1}}(z) \, .
\label{eq:SPO6}
\end{equation}
Thus, the factorization operators are completely determined once we compute the \textit{seed function} $u_{\epsilon_{1}}(z)$, which turns out to be an eigenfunction of the stationary oscillator with eigenvalue $\epsilon_{1}$, it can be either a physical or a nonphysical solution. Its explicit form will be discussed in the following sections.

Now, we construct a new operator $I_{1}(t)$ by inverting the order of the factorization in~\eqref{eq:SPO1}, after some algebra we obtain
\begin{equation}
I_{1}(t):=B_{1}(t)B_{1}^{\dagger}(t)+\epsilon_{1}=I_{0}(t)+2\sigma\frac{\partial W_{1}}{\partial x}=I_{0}(t)-2\frac{\partial^{2}}{\partial z^{2}}\ln u_{\epsilon_{1}}(z) \, .
\label{eq:SPO7}
\end{equation}
It is clear that $I_{1}(t)$ is not in general a constant of motion of $H_{0}(t)$, but we can find a time-dependent Hamiltonian $H_{1}(t)$ for which $I_{1}(t)$ is the respective invariant. To this end we introduce the ansatz $H_{1}(t)=H_{0}(t)+G(t)\mathfrak{F}(z)$, with $G(t)$ and $\mathfrak{F}(z)$ to be determined from the condition
\begin{equation}
\frac{dI_{1}(t)}{dt}=i[H_{1}(t),I_{1}(t)]+\frac{\partial I_{1}(t)}{\partial t}=0 \, .
\label{eq:SPO8}
\end{equation}
Straightforward calculation shows that $H_{1}(t)$ is given, in coordinate representation, as (see App.~\ref{sec:APPB} for details)
\begin{equation}
\begin{aligned}
& H_{1}(t)=-\frac{\partial^{2}}{\partial x^{2}}+V_{1}(x,t) \, , \quad V_{1}(x,t)=V_{0}(x,t)-\frac{2}{\sigma^{2}}\frac{\partial^{2}}{\partial z^{2}}\ln u_{\epsilon_{1}}(z) \, ,
\end{aligned}
\label{eq:SPO9}
\end{equation}
where
\begin{equation}
V_{0}(x,t)= \Omega^{2}(t)x^{2}+F(t)x \, .
\label{eq:POTPO}
\end{equation}
Thus, the solutions of the Schr\"odinger equation
\begin{equation}
i\frac{\partial}{\partial t}\psi^{(1)}=H_{1}(t)\psi^{(1)} \, ,
\label{eq:SPO91}
\end{equation}
are computed following the discussion of Sec.~\ref{sec:PO}. Indeed, we first solve the spectral problem
\begin{equation}
I_{1}(t)\varphi^{(1)}_{n}(x,t)=\lambda_{n}^{(1)}\varphi_{n}^{(1)}(x,t) \, ,
\label{eq:SPO10}
\end{equation}
with $\lambda_{n}^{(1)}$ and $\varphi_{n}^{(1)}(x,t)$ being the eigenvalues and eigenfunctions of $I_{1}(t)$=\eqref{eq:SPO7}, respectively. Such a task is achieved by using the \textit{intertwining relationships} between $I_{0}$ and $I_{1}$, constructed from~\eqref{eq:SPO1} and~\eqref{eq:SPO7} as
\begin{subequations}
\begin{equation}
B_{1}I_{0}(t)=I_{1}(t)B_{1}\, ,
\label{eq:SPO10a}
\end{equation}
\begin{equation}
B_{1}^{\dagger}I_{1}(t)=I_{0}(t)B_{1}^{\dagger} \, .
\label{eq:SPO10b}
\end{equation}
\end{subequations}
Thus, it is clear that the respective action of~\eqref{eq:SPO10a} and~\eqref{eq:SPO10b} on the eigenfunctions $\varphi_{n}^{(0)}(x,t)$ and $\varphi_{n}^{(1)}(x,t)$ leads to
\begin{subequations}
\begin{equation}
I_{1}(t)\left(B_{1}\vert\varphi^{(0)}_{n}(t)\rangle\right)=\lambda_{n}^{(0)}\left(B_{1}\vert\varphi_{n}^{(0)}(t)\rangle\right) \, ,
\label{eq:SPO11a}
\end{equation}
\begin{equation}
I_{0}(t)\left(B^{\dagger}_{1}\vert\varphi^{(1)}_{n}(t)\rangle\right)=\lambda_{n}^{(1)}\left(B^{\dagger}_{1}\vert\varphi_{n}^{(1)}(t)\rangle\right) \, ,
\label{eq:SPO11b}
\end{equation}
\end{subequations}
The operator $B_{1}$ maps $\varphi_{n}^{(0)}$ into an eigenfunction of $I_{1}(t)$ with eigenvalue $\lambda_{n}^{(0)}$, whereas $B_{1}^{\dagger}$ reverse the mapping. From~\eqref{eq:SPO11a}-\eqref{eq:SPO11b}, we construct the orthonormal set $\mathcal{S}^{(1)}=\{ \varphi^{(1)}_{\theta_{n}} \}_{n=0}^{\infty}$, with elements
\begin{equation}
\varphi_{\theta_{n}}^{(1)}(x,t)=\frac{1}{\sqrt{\lambda^{(0)}_{n}-\epsilon_{1}}}B_{1}\varphi_{n}^{(0)}(x,t) \, , \quad n=0,1,\cdots \, ,
\label{eq:SPO12}
\end{equation}
where $\theta_{n}$ is an ordering function that dictates how to arrange the eigenfunctions according to their eigenvalues or, equivalently, to their number of nodes. The normalization factor has been determined from the properties of the intertwining relations and the factorization~\eqref{eq:SPO1}. Before discussing the form of $\theta_{n}$ we inspect the completeness of the set $\mathcal{S}^{(1)}$. 

Let us suppose that there is an eigenfunction $\varphi_{\epsilon_{1}}^{(1)}$, henceforth called the \textit{missing state}, orthogonal to $\varphi_{\theta_n}^{(1)}$, for every $n=0,1,\cdots$. If the missing state is the trivial solution, $\varphi_{\epsilon_{1}}^{(1)}=0$, we say that $\mathcal{S}^{(1)}$ is already a complete set of solutions. On the other hand, if the missing state has finite-norm, it must be added to the set of solution, that is, $\mathcal{S}_{c}^{(1)}=\{ \varphi_{\epsilon_{1}}^{(1)}(x,t)\}\cup\mathcal{S}^{(1)}$. The orthogonality condition of the missing state with respect to each $\varphi_{\theta_{n}}^{(1)}$=\eqref{eq:SPO12} implies
\begin{equation}
\langle \varphi_{\epsilon_{1}}^{(1)}(t)\vert B_{1}\vert\varphi_{n}^{(0)}(t)\rangle = 0 \, , \quad n=0,1,\cdots \, ,
\label{eq:SPO121}
\end{equation}
from where we obtain the condition $B_{1}^{\dagger}\vert\varphi_{\epsilon_{1}}^{(1)}(t)\rangle=0$. After some algebra we arrive to
\begin{equation}
\varphi_{\epsilon_{1}}^{(1)}(x,t)=\mathcal{N}_{\epsilon_{1}}\frac{e^{i\frac{\dot{\sigma}}{4\sigma}(x+\gamma)^{2}-i\frac{\dot{\gamma}}{2}x}}{\sqrt{\sigma} u_{1}(z(x,t))} \, ,
\label{eq:SPO13}
\end{equation}
with $\mathcal{N}_{\epsilon_{1}}$ the normalization constant. From $B_{1}^{\dagger}\vert\varphi_{\epsilon_{1}}^{(1)}(t)\rangle=0$ and the factorization~\eqref{eq:SPO7} it follows that the missing state corresponds to the eigenvalue $\lambda^{(1)}=\epsilon_{1}$. Moreover, if $\vert\mathcal{N}_{\epsilon_{1}}\vert^{2}\neq 0$, the ordering function takes the form $\theta_{n}=n+1$ and the ordered eigenfunctions and eigenvalues are respectively given by
\begin{equation}
\varphi_{0}^{(1)}(x,t)=\varphi_{\epsilon_{1}}^{(1)}(x,t) \, , \quad \varphi_{n+1}^{(1)}(x,t)=\frac{1}{\sqrt{\lambda_{n}^{(0)}-\epsilon_{1}}} \, B_{1}\varphi_{n}^{(0)}(x,t) \, ,
\label{eq:SPO14}
\end{equation}
\begin{equation}
\lambda_{0}^{(1)}=\epsilon_{1} \, , \quad \lambda_{n+1}^{(1)}=\lambda_{n}^{(0)}=2n+1 \, , \quad n=0,1,\cdots \, .
\label{eq:SPO14-1}
\end{equation}

Now, the solutions of the Schr\"odinger equation~\eqref{eq:SPO91} are determined by adding the complex-phase $\chi_{n}^{(1)}(t)$ to $\varphi_{n}^{(1)}(x,t)$ as 
\begin{equation}
\psi_{n}^{(1)}(x,t)=e^{i\chi_{n}^{(1)}(t)}\varphi_{n}^{(1)}(x,t) \, ,
\label{eq:SPO17}
\end{equation}
where
\begin{equation}
\dot{\chi}^{(1)}(t)=\langle\varphi_{n}^{(1)}(t)\vert i\frac{\partial}{\partial t}-H_{1}(t)\vert\varphi_{n}^{(1)}(t)\rangle \, .
\label{eq:SPO15}
\end{equation}
After some calculations we get (see App.\ref{sec:APPC})
\begin{equation}
\chi^{(1)}_{n}(t)=-\lambda_{n}^{(1)}\int^{t}\frac{dt'}{\sigma^{2}(t')}-\frac{\gamma\dot{\gamma}}{4}+\frac{1}{2}\int^{t}dt'\gamma(t')F(t') \, .
\label{eq:SPO16}
\end{equation}

We thus have constructed a new time-dependent Hamiltonian $H_{1}(t)$, together with the respective complete set of solutions $\{ \psi^{(1)}_{n}(x,t) \}_{n=0}^{\infty}$. A summary of the method is depicted in the diagram of Fig.\ref{fig:F0}.

\begin{figure}
\centering
\begin{tikzpicture}
  \matrix (m) [matrix of math nodes,row sep=3em,column sep=4em,minimum width=2em]
  {
     i\frac{\partial}{\partial t}\psi^{(0)}=H_{0}(t)\psi^{(0)}
     & i\frac{\partial}{\partial t}\psi^{(1)}=H_{1}(t)\psi^{(1)} \\
     I_{0}(t)\varphi_{n}^{(0)}=\lambda_{n}^{(0)}\varphi_{n}^{(0)} 
     & I_{1}(t)\varphi_{n}^{(1)}=\lambda_{n}^{(1)}\varphi_{n}^{(1)} \\
     I_{0}(t)=B^{\dagger}_{1}B_{1}+\epsilon_{1} 
     & I_{1}(t)=B_{1}B^{\dagger}_{1}+\epsilon_{1}  \\};
  \path[-stealth]
    (m-1-1) edge [double] node [left] {$\psi_{n}^{(0)}=e^{i\chi^{(0)}_{n}(t)}\varphi^{(0)}_{n}$} (m-2-1)
    (m-2-1) edge [double] node [left] {} (m-3-1)
    (m-3-1) edge [double] node [below] {} (m-3-2)
    (m-3-2) edge [double] node [right] {} (m-2-2)
    (m-2-2) edge [double] node [right] {$\psi_{n}^{(1)}=e^{i\chi^{(1)}_{n}(t)}\varphi^{(1)}_{n}$} (m-1-2);
\end{tikzpicture}
\caption{Schematic procedure of the factorization method applied to the invariant operator of the parametric oscillator.}
\label{fig:F0}
\end{figure}

\subsubsection{Shape invariant case}
\label{subsubsec:SHIN}
As a special case we have $\epsilon_{1}=1$, where $u_{\epsilon_{1}}=e^{-z^{2}/2}$ is a particular solution of~\eqref{eq:SPO6}. This leads to $W_{1}=z$ and $\mathcal{F}_{1}=0$. Thus, the factorization operators $B_{1}$ and $B_{1}^{\dagger}$ reduce respectively to the ladder operators $A$ and $A^{\dagger}$ given in \eqref{eq:PO10}. The new quantum invariant and the Hamiltonian take the form
\begin{equation}
\end{equation}
respectively. The invariant operator $I_{1}(t)$ is just the initial one displaced by two units. Thus, we can identify $I_{1}(t)$ with the class of shape invariant operators related to $I_{0}(t)$ (see~\cite{Kha93,Car00} for examples in the stationary case). Moreover, $I_{1}(t)$ is a constant of motion of both $H_{0}(t)$ and $H_{1}(t)$, that is,
\begin{equation}
\frac{d I_{1}(t)}{dt}=i[H_{0}(t),I_{1}(t)]+\frac{\partial I_{1}}{\partial t}=i[H_{1},I_{1}(t)]+\frac{\partial I_{1}(t)}{\partial t}= 0 \, .
\end{equation}
Therefore, $H_{0}(t)$ and $H_{1}(t)$ must lead to equivalent solutions. To prove the latter, let us consider $\psi^{(0)}$ and $\psi^{(1)}$ as the respective solutions of the Schr\"odinger equations
\begin{equation}
i\frac{\partial}{\partial t}\psi^{(0)}=H_{0}(t)\psi^{(0)} \, , \quad i\frac{\partial}{\partial t}\psi^{(1)}=H_{1}(t)\psi^{(1)}=\left(H_{0}(t)+\frac{2}{\sigma^{2}}\right)\psi^{(1)} \, ,
\label{eq:SHIN3}
\end{equation}
and suppose that both solutions are related through a time-dependent factor, $\psi^{(1)}=g(t)\psi^{(0)}$. After substituting the latter in the Schr\"odinger equation for $H_{1}(t)$ in~\eqref{eq:SHIN3} we get
\begin{equation}
\psi^{(1)}(x,t)=e^{-2i\int^{t}dt' \, \sigma^{-2}(t')}\psi^{(0)}(x,t) \, .
\end{equation}
Then, the solutions differ only by a global complex-phase and both $H_{0}(t)$ and $H_{1}(t)$ are equivalent. This also explains why $\mathcal{V}(t)=0$ in~\eqref{eq:PO1} is as general as the case $\mathcal{V}(t)\neq 0$.

\subsection{Two-step factorization}
\label{subsec:2SF}
The factorization developed so far can be iterated as many times as needed. In each iteration we construct a new invariant operator and the time-dependent Hamiltonian related to it. For simplicity, we consider the \textit{two-step factorization}. Higher orders will be obtained in complete analogy. 

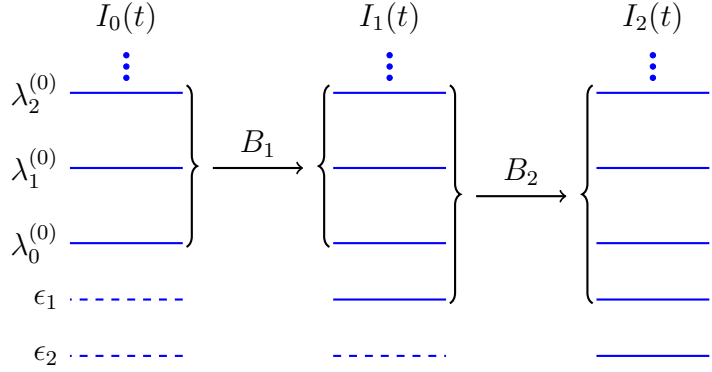
\begin{figure}
\centering
\begin{tikzpicture}
\draw (0.75,3) node {$I_{0}(t)$};
\draw (4.25,3) node {$I_{1}(t)$};
\draw (7.75,3) node {$I_{2}(t)$};
\filldraw[blue] (0.75,2.2) circle (1pt);
\filldraw[blue] (0.75,2.35) circle (1pt);
\filldraw[blue] (0.75,2.5) circle (1pt);
\filldraw[blue] (4.25,2.2) circle (1pt);
\filldraw[blue] (4.25,2.35) circle (1pt);
\filldraw[blue] (4.25,2.5) circle (1pt);
\filldraw[blue] (7.75,2.2) circle (1pt);
\filldraw[blue] (7.75,2.35) circle (1pt);
\filldraw[blue] (7.75,2.5) circle (1pt);
\draw[blue,thick] (1.5,2)--(0,2) node[anchor=east,text=black] {$\lambda_{2}^{(0)}$};
\draw[blue,thick] (1.5,1)--(0,1) node[anchor=east,text=black] {$\lambda_{1}^{(0)}$};
\draw[blue,thick] (1.5,0)--(0,0) node[anchor=east,text=black] {$\lambda_{0}^{(0)}$};
\draw[blue,thick,dashed] (1.5,-0.75)--(0,-0.75) node[anchor=east,text=black] {$\epsilon_{1}$};
\draw[blue,thick,dashed] (1.5,-1.5)--(0,-1.5) node[anchor=east,text=black] {$\epsilon_{2}$};
\draw[blue,thick] (3.5,2)--(5,2);
\draw[blue,thick] (3.5,1)--(5,1);
\draw[blue,thick] (3.5,0)--(5,0);
\draw[blue,thick] (3.5,-0.75)--(5,-0.75);
\draw[blue,thick,dashed] (3.5,-1.5)--(5,-1.5);
\draw[blue,thick] (7,2)--(8.5,2);
\draw[blue,thick] (7,1)--(8.5,1);
\draw[blue,thick] (7,0)--(8.5,0);
\draw[blue,thick] (7,-0.75)--(8.5,-0.75);
\draw[blue,thick] (7,-1.5)--(8.5,-1.5);
\draw[thick,decorate,decoration={brace,amplitude=4pt}]
(1.55,2.1)--(1.55,-0.05);
\draw[thick,decorate,decoration={brace,amplitude=4pt,mirror}]
(3.45,2.1)--(3.45,-0.05);
\draw[thick,decorate,decoration={brace,amplitude=4pt}]
(5.05,2.1)--(5.05,-0.8);
\draw[thick,decorate,decoration={brace,amplitude=4pt,mirror}]
(6.95,2.1)--(6.95,-0.8);
\draw[thick,->] (1.9,1)--(3.1,1) node[midway,anchor=south] {$B_{1}$};
\draw[thick,->] (5.4,0.625)--(6.6,0.625) node[midway,anchor=south] {$B_{2}$};
\end{tikzpicture}
\caption{Eigenvalues associated with the quantum invariant of the parametric oscillator and the invariants obtained through the one-step and two-step factorizations. Solid and dashed lines depict the physical and non-physical eigenvalues, respectively.}
\label{fig:2S}
\end{figure}

Following the construction of Sec.~\ref{subsec:1SF}, we introduce an alternative couple of operators
\begin{equation}
\begin{aligned}
B_{2} := A + \mathcal{F}_{2} \, , \quad  B^{\dagger}_{2}:= A^{\dagger} + \mathcal{F}_{2} \, , \quad \mathcal{F}_{2}=-z+W_{2} \, , \quad z=\frac{x+\gamma}{\sigma} ,
\end{aligned}
\label{eq:MS1}
\end{equation}
where $\mathcal{F}_{2}(x,t)$ and $W_{2}(z)$ are real-valued functions. The operators $B_{2}$ and $B_{2}^{\dagger}$ factorize the previously generated invariant operator $I_{1}(t)$=\eqref{eq:SPO7} as
\begin{equation}
I_{1}(t)=B_{2}^{\dagger}B_{2}+\epsilon_{2} \, , \quad \epsilon_{2}<\lambda_{0}^{(0)} \, .
\label{eq:MS2}
\end{equation}
After some algebra we recover again a Riccati equation for $W_{2}$, solved through the linear equation
\begin{equation}
-\frac{\partial^{2}v_{\epsilon_{2}}}{\partial z^{2}}+\left(z^{2}+2\frac{\partial W_{1}}{\partial z} \right)v_{\epsilon_{2}}=\epsilon_{2} v_{\epsilon_{2}} \, , \quad W_{2}=-\frac{1}{v_{\epsilon_{2}}}\frac{\partial v_{\epsilon_{2}}}{\partial z} \, ,
\label{eq:MS3}
\end{equation}
with $v_{\epsilon_{2}}(z)$ the seed function of the two-step factorization. Notice that $v_{\epsilon_{2}}$ solves an eigenvalue problem associated with a deformed oscillator potential, and it can be rewritten in terms of the respective seed functions of the one-step case~\eqref{eq:SPO6} as
\begin{equation}
v_{\epsilon_{2}}=\frac{\mathcal{W}(\epsilon_{1},\epsilon_{2})}{u_{\epsilon_{1}}} \, , \quad \mathcal{W}(\epsilon_{1},\epsilon_{2})=u_{\epsilon_{1}}\frac{\partial u_{\epsilon_{2}}}{\partial z}-u_{\epsilon_{2}}\frac{\partial u_{\epsilon_{1}}}{\partial z} \, ,
\label{eq:MS3-1}
\end{equation}
where $u_{\epsilon_{2}}(z)$ solves the harmonic oscillator eigenvalue equation~\eqref{eq:SPO6} with eigenvalue $\epsilon_{2}$. We thus rewrite $W_{2}(z)$ in terms of the seed functions $u_{\epsilon_{1}}$ and $u_{\epsilon_{2}}$ as
\begin{equation}
W_{2}(x,t)=-\frac{\partial}{\partial z}\ln \mathcal{W}(\epsilon_{1},\epsilon_{2})-W_{1}(x,t) \, , 
\label{eq:MS3-2}
\end{equation}
with $W_{1}(x,t)$ given in~\eqref{eq:SPO6}. Now, the inverted factorization of~\eqref{eq:MS2} leads to a new invariant operator 
\begin{equation}
I_{2}(t)=B_{2}B_{2}^{\dagger}+\epsilon_{2}=I_{1}+2\sigma\frac{\partial}{\partial x}W_{2}=I_{0}+2\frac{\partial}{\partial z}\mathcal{W}(\epsilon_{1},\epsilon_{2}) \, .
\label{eq:MS4}
\end{equation}
In a similar way to the one-step case, the new time-dependent Hamiltonian is given by (see App.~\ref{sec:APPB})
\begin{equation}
H_{2}(t)=-\frac{\partial^{2}}{\partial x^{2}}+V_{2}(x,t) \, ,
\label{eq:MS401}
\end{equation}
where the new time-dependent potential is
\begin{equation}
V_{2}(x,t)=V_{1}(x,t)+\frac{2}{\sigma^{2}}\frac{\partial W_{2}}{\partial z}=V_{0}(x,t)-\frac{2}{\sigma^{2}}\frac{\partial^{2}}{\partial z^{2}}\ln\mathcal{W}(\epsilon_{1},\epsilon_{2}) \, ,
\label{eq:MS5}
\end{equation}
with $V_{0}(x,t)$ given in~\eqref{eq:POTPO}. Then, the spectral information of the eigenvalue problem
\begin{equation}
I_{2}(t)\varphi_{n}^{(2)}(x,t)=\lambda_{n}^{(2)}\varphi_{n}^{(2)}(x,t) \, ,
\label{eq:MS5-1}
\end{equation}
is determined from $I_{1}$, which has been already solved in the previous section. With the use of both the factorization~\eqref{eq:MS2} and~\eqref{eq:MS4}, along with the intertwining relations of the one-step case~\eqref{eq:SPO10a}-\eqref{eq:SPO10b}, we arrive to relationships of the form
\begin{equation}
\left\{ \begin{aligned} &I_{2}B_{2}=B_{2}I_{1} \\ &I_{1}B_{2}^{\dagger}=B^{\dagger}I_{2} \end{aligned} \right. \, , \qquad \left\{ \begin{aligned} &I_{2}B_{2}B_{1}=B_{2}B_{1}I_{0} \\ &I_{0}B_{1}^{\dagger}B_{2}^{\dagger}=B_{1}^{\dagger}B_{2}^{\dagger}I_{2} \end{aligned} \right. \, .
\label{eq:MS6}
\end{equation}
The first set of intertwining relations in~\eqref{eq:MS6} allows to connect the spectral information of $I_{2}$ with that of $I_{1}$, see Fig.~\ref{fig:2S}. In turn, the second set of relations connects $I_{2}$ directly to the initial quantum invariant $I_{0}$. Thus, $B_{2}$ maps the eigenfunctions of $I_{1}$ into eigenfunctions of $I_{2}$, and $B_{2}B_{1}$ maps the eigenfunctions of $I_{0}$ into the respective ones of $I_{2}$. Additionally, in analogy to the discussion of the missing state of Sec.\eqref{sec:FAC}, there is an eigenfunction $\varphi_{\epsilon_{2}}$ associated with the eigenvalue $\epsilon_{2}$ which is not obtained from the previous mappings. Therefore, we construct the complete set of normalized eigenfunctions $\mathcal{S}^{(2)}_{c}=\{ \varphi_{n}^{(2)} \}_{n=0}^{\infty}$ as
\begin{equation}
\begin{aligned}
& \varphi_{n+2}^{(2)}=\frac{B_{2}\varphi_{n+1}^{(1)}}{\sqrt{\lambda_{n+1}^{(1)}-\epsilon_{2}}}=\frac{B_{2}B_{1}\varphi_{n}^{(0)}}{\sqrt{\left(\lambda_{n}^{(0)}-\epsilon_{2} \right)\left( \lambda_{n}^{(0)}-\epsilon_{1} \right)}} \, , \quad n=0,1,\cdots \, ,\\
& \varphi_{1}^{(2)}=\frac{B_{2}\varphi_{0}^{(1)}}{\sqrt{\lambda_{0}^{(1)}-\epsilon_{2}}} = \frac{\mathcal{N}_{\epsilon_{1}}}{\sqrt{\epsilon_{1}-\epsilon_{2}}} \, \frac{e^{i\left( \frac{\sigma\dot{\sigma}}{4}z^{2}-\frac{\sigma\dot{\gamma}}{2}z \right)}}{\sqrt{\sigma}}\frac{u_{\epsilon_{2}}}{\mathcal{W}(\epsilon_{1},\epsilon_{2})} \, , \\
& \varphi_{0}^{(2)}=\mathcal{N}_{\epsilon_{2}} \, \frac{e^{i\left( \frac{\sigma\dot{\sigma}}{4}z^{2}-\frac{\sigma\dot{\gamma}}{2}z \right)}}{\sqrt{\sigma}}\frac{u_{\epsilon_{1}}}{\mathcal{W}(\epsilon_{1},\epsilon_{2})} \, ,
\end{aligned}
\label{eq:MS7}
\end{equation}
where $\mathcal{N}_{\epsilon_{1}}$ is the same normalization constant of the missing state $\varphi_{0}^{(1)}$~\eqref{eq:SPO13} and $\mathcal{N}_{\epsilon_{2}}$ is the normalization constant of the additional missing state of $I_{2}(t)$. Notice that $\varphi_{n}^{(2)}$ has been expressed in terms of the eigenfunctions of $I_{0}(t)$ and the seed functions $u_{\epsilon_{1,2}}$, which are all already known. To complete the spectral information, the eigenvalues for~\eqref{eq:MS5-1} are 
\begin{equation}
\lambda^{(2)}_{n+2}=\lambda_{n}^{(0)}=2n+1 \, , \quad \lambda_{1}^{(2)}=\epsilon_{1} \, , \quad \lambda_{0}^{(2)}=\epsilon_{2} \, .
\label{eq:MS8}
\end{equation}
Notice that, to obtain a well-behaved potential and finite-norm solutions it is required that $\mathcal{W}(\epsilon_{1},\epsilon_{2})$ be a nodeless function for $x\in\mathbb{R}$. For the one-step factorization we have already fixed $u_{\epsilon_{1}}$ to be nodeless. Thus, the seed function $u_{\epsilon_{2}}$ might be or not a function with nodes and it must be fixed with additional caution. A discussion on that matter is provided in the next section.

From~\eqref{eq:MS5}, we obtain the solutions to the Schr\"odinger equation
\begin{equation}
i\frac{\partial\psi_{n}^{(2)}}{\partial t}=H_{2}(t)\psi_{n}^{(2)} \, , 
\label{eq:MS9}
\end{equation}
by multiplying the time-dependent complex-phase $\chi_{n}^{(2)}$ (see App.\ref{sec:APPC} for details) to the eigenfunctions $\varphi_{n}^{(2)}$,
\begin{equation}
\begin{aligned}
& \psi_{n}^{(2)}(x,t)=e^{i\chi_{n}^{(2)}(t)}\varphi_{n}^{(2)}(x,t) \, , \\
& \chi_{n}^{(2)}(t)=-\frac{\gamma\dot{\gamma}}{4}-\lambda_{n}^{(2)}\int^{t}\frac{dt'}{\sigma^{2}(t')}+\frac{1}{2}\int^{t}dt'\gamma(t')F(t') \, .
\end{aligned}
\label{eq:MS10}
\end{equation}
The integral of $\sigma^{-2}$ in~\eqref{eq:MS10} has been already computed in~\eqref{eq:PO6-1}. It is just required to compute the last integral~\eqref{eq:MS10}, although, it has to be done once we specify the form of $F(t)$.

\section{Time-dependent rational extensions of the parametric oscillator}
\label{sec:REX}
The construction of exactly solvable and almost-isospectral Hamiltonians related with the harmonic oscillator is well documented. Among those models, the rational extensions play an important role since those lead to a new family of Hamiltonians with eigenfunctions that belong to the class of exceptional Hermite polynomials~\cite{Gom14,Mar13,Hof19}. In this section we explore the construction presented in previous sections to construct the time-dependent counterparts of rational extensions associated with the parametric oscillator. 

\subsection{One-step rational extension}
\label{subsec:OSRT}
In order to construct the family of rational potentials associated to the one-step factorization, it is required that $W_{1}$ become a rational function of $x$~\cite{Mar13}. Moreover, the seed function must also be a nodeless function to avoid singularities. The most general form of the seed function, computed from~\eqref{eq:SPO6}, takes the form
\begin{equation}
u_{\epsilon_{1}}= \eta_{0} e^{-z^{2}/2} \, {}_{1}F_{1}\left(\frac{1-\epsilon_{1}}{4},\frac{1}{2};z^{2}\right)+\eta_{1}e^{-z^{2}/2}z \,{}_{1}F_{1}\left(\frac{3-\epsilon_{1}}{4},\frac{3}{2};z^{2}\right) \, ,
\label{eq:REX01}
\end{equation}
where $\epsilon_{1}$, $\eta_{0}$, $\eta_{1}$ are real constants and ${}_{1}F_{1}(\cdot,\cdot;z)$ stands for the \textit{confluent hypergeometric function}~\cite{Olv10}. From~\eqref{eq:REX01} we realize that $u_{\epsilon_{1}}$ becomes proportional to the even Hermite polynomials $\mathtt{H}_{2m}(z)$ for $\{ \epsilon_{1}=4m+1,\eta_{1}=0 \}$ and the odd Hermite polynomials $\mathtt{H}_{2m+1}(z)$ for $\{ \epsilon_{1}=4m+3, \eta_{0}=0\}$, where $m=0,1,\cdots$. The $n-$th Hermite polynomial has exactly $n$ zeros, and thus $\mathtt{H}_{0}(z)=1$ is the only well behaved seed function. However, from Sec.\ref{subsubsec:SHIN}, such a solution leads to the family of shape invariant potentials, which are not relevant for the present work. A second family of polynomial solutions is obtained from~\eqref{eq:REX01} with aid of the Kummer transformation~\cite{Olv10}, leading to
\begin{equation}
u_{\epsilon_{1}}(z)=\eta_{0} e^{z^{2}/2} \, {}_{1}F_{1}\left(\frac{1+\epsilon_{1}}{4},\frac{1}{2};-z^{2}\right)+\eta_{1}e^{z^{2}/2}z \,{}_{1}F_{1}\left(\frac{3+\epsilon_{1}}{4},\frac{3}{2};-z^{2}\right) \, .
\label{eq:REX02}
\end{equation}
From the latter, even and odd polynomials are determined from the conditions $\{\epsilon_{1}=-4m-1,\eta_{1}=0\}$ and $\{\epsilon_{1}=-4m-3,\eta_{0}=0\}$, respectively, with $m=0,1,\cdots$. After some computation we get, up to a proportional constant,
\begin{equation}
u_{\epsilon_{1}=-2m-1}(z)=e^{\frac{1}{2}z^{2}}\mathcal{H}_{m}(z) \, , \quad m=0,1,\cdots \, ,
\label{eq:REX3}
\end{equation}
where $\mathcal{H}_{m}(z)$ stands for the \textit{pseudo-Hermite polynomials}~\cite{Abr72} given by
\begin{equation}
\mathcal{H}_{m}(z)=(-i)^{m}\mathtt{H}_{m}(iz)=m!\sum_{p=0}^{\lfloor m/2 \rfloor}\frac{(2z)^{m-2p}}{p!(m-2p)!} \, ,
\label{eq:REX4}
\end{equation}
with $\lfloor \cdot \rfloor$ the \textit{floor function}~\cite{Olv10}. The pseudo-Hermite polynomials emerge naturally in the construction of the family of \textit{exceptional Hermite polynomials}~\cite{Mar13}, resulting from the rational extensions of the harmonic oscillator. 

From~\eqref{eq:REX3} we can notice that the condition $\epsilon_{1}<\lambda^{(0)}_{0}=1$ is automatically fulfilled.  Additionally, from~\eqref{eq:REX4} it is clear that $\mathcal{H}_{2m+1}(z)$ has one node at the origin, whereas $\mathcal{H}_{2m}(z)$ is nodeless. Thus, we consider for the one-step construction the even solutions of~\eqref{eq:REX4}. In this form we obtain the new time-dependent potential
\begin{equation}
\begin{aligned}
& V_{1}(x,t)=\Omega^{2}(t)x^{2}+F(t)x-\frac{2}{\sigma^{2}}\left(1+\frac{\partial^{2}}{\partial z^{2}}\ln\mathcal{H}_{m}(z) \right) \, , \quad m=2,4,\cdots \, .
\end{aligned}
\label{eq:REX6}
\end{equation}
The eigenfunctions of the invariant operator $I_{1}(t)$ are computed from~\eqref{eq:SPO14} which, after some algebra, are given by
\begin{equation}
\begin{aligned}
& \varphi_{0}^{(1)}(x,t)=\mathcal{N}_{0}^{(m)}\frac{\exp\left[\left(-\frac{1}{2\sigma^{2}}+i\frac{\dot{\sigma}}{4\sigma} \right)(x+\gamma)^{2}-i\frac{\dot{\gamma}}{2}x\right]}{\sqrt{\sigma}\mathcal{H}_{m}(z)} \, , \\
& \varphi_{n+1}^{(1)}(x,t)=\mathcal{N}_{n+1}^{(m)}\frac{\exp\left[\left(-\frac{1}{2\sigma^{2}}+i\frac{\dot{\sigma}}{4\sigma} \right)(x+\gamma)^{2}-i\frac{\dot{\gamma}}{2}x\right]}{\sqrt{\sigma}\mathcal{H}_{m}(z)}P_{n}^{(m)}(z) \, , \\ 
& P_{n}^{(m)}(z)=-\mathcal{H}_{m}(z)\mathtt{H}_{n+1}(z)-2m\mathcal{H}_{m-1}(z)\mathtt{H}_{n}(z) \, , \quad m=2,4,\cdots \, , \quad n=0,1,\cdots ,
\end{aligned}
\label{eq:REX10}
\end{equation}
where 
\begin{equation}
\mathcal{N}_{n+1}^{(m)}=\left( 2^{n+1}n!(n+m+1)\sqrt{\pi}\right)^{-1/2} \, , \quad \mathcal{N}_{0}^{(m)}=\left( \frac{2^{m}m!}{\sqrt{\pi}} \right)^{1/2} \, .
\label{eq:REX10-1}
\end{equation}
The normalization constants $\mathcal{N}_{n+1}^{(m)}$ were determined from~\eqref{eq:SPO14} for $n=0,1,\cdots$. In turn, $\mathcal{N}_{0}^{(m)}$ has to be computed explicitly. From the definition of inner product~\eqref{eq:PO72} and the reparametrization $z(x,t)$ it is easy to show that $\mathcal{N}_{0}^{(m)}$ is determined from the same relation of the stationary case. Therefore, the normalization constant of the rational extension of the harmonic oscillator, reported in~\cite{Mar13}, was used. 

The respective solutions to the Schr\"odinger equation are computed from the relation~\eqref{eq:SPO17}. 

To illustrate the form of the potentials generated from the one-step rational extensions, let us consider $m=4$ and the rest of parameters arbitrary. After some calculation we obtain
\begin{equation}
V_{1}(x,t)=\Omega^{2}(t)x^{2}+F(t)x-\frac{2}{\sigma^{2}}\left[1-8\left(\frac{8z^{6}+12z^4+18z^2-9}{16z^{8}+96z^{6}+168z^{4}+72z^{2}+9} \right) \right] \, ,
\label{eq:REX11}
\end{equation}
where $z(x,t)=(x+\gamma)/\sigma$, with $\sigma$ and $\gamma$ to be determined once $\Omega^{2}(t)$ and $F(t)$ are specified. Notice that $V_{1}(x,t)$ includes a term $2/\sigma^{2}$ that depends only on time, as discussed in Sec.~\ref{subsubsec:SHIN}, it can be eliminated from the potential by adding the appropriate global phase to the solutions. Nevertheless, for the sake of simplicity we preserve such a term.

\subsection{Two-step rational extension}
\label{subsec:TSRT}
For the two-step factorization we have to construct a nodeless function $\mathcal{W}(\epsilon_{1},\epsilon_{2})$ which, with the use of the pseudo-Hermite polynomials, takes the form
\begin{equation}
\begin{aligned}
& \mathcal{W}(m_{1},m_{2})\equiv\mathcal{W}(\epsilon_{1}=m_{1},\epsilon_{2}=m_{2})=e^{z^2}g_{m_{1},m_{2}}(z) \, ,  \\
& g_{m_{1},m_{2}}(z)=2m_{2}\mathcal{H}_{m_{1}}(z)\mathcal{H}_{m_{2}-1}(z)-2m_{1}\mathcal{H}_{m_{1}-1}(z)\mathcal{H}_{m_{2}}(z) \, .
\end{aligned}
\label{eq:REX13}
\end{equation}
From the form of $\mathcal{H}_{m}$~\eqref{eq:REX4}, we can see that $g_{m_{1},m_{2}}(z)$ admits one node whenever $m_{1}$ and $m_{2}$ are both even or both odd. On the other hand, for $m_{1}$ ($m_{2}$) even and $m_{2}$ ($m_{1}$) odd we obtain a nodeless function. Notice that $\mathcal{W}(m_{1},m_{2})=-\mathcal{W}(m_{2},m_{1})$ and the new potentials, together with the new eigenfunctions, are invariant under such a change. The latter also implies that $m_{1}$ and $m_{2}$ are not necessarily ordered, that is, either $m_{1}<m_{2}$ or $m_{2}<m_{1}$. However, for the sake of simplicity, we consider $m_{1}<m_{2}$ in the rest of the text. 

From~\eqref{eq:MS5} we obtain the two-step time-dependent rational extension 
\begin{equation}
V_{2}(x,t)=\Omega^{2}(t)x^{2}+F(t)x-\frac{2}{\sigma^{2}}\left(2+\frac{\partial^{2}}{\partial z^{2}}\ln g_{m_{1},m_{2}}(z) \right) \, ,
\label{eq:REX14}
\end{equation}
where the respective eigenfunctions of $I_{2}(t)$ are given by
\begin{equation}
\begin{aligned}
& \varphi_{n+2}^{(2)}(x,t)=\mathcal{N}_{n+2}^{(m_{1},m_{2})}\frac{e^{i\left( \frac{\sigma\dot{\sigma}}{4}z^{2}-\frac{\sigma\dot{\gamma}}{2}z \right)}}{\sqrt{\sigma}}\frac{e^{-z^2/2}}{g_{m_{1},m_{2}}(z)}P^{(m_{1},m_{2})}_{n}(z) \, , \\
& \varphi_{1}^{(2)}(x,t)=\mathcal{N}_{1}^{(m_{1},m_{2})} \frac{e^{i\left( \frac{\sigma\dot{\sigma}}{4}z^{2}-\frac{\sigma\dot{\gamma}}{2}z \right)}}{\sqrt{\sigma}} \frac{e^{-z^2/2}}{g_{m_{1},m_{2}}(z)}\mathcal{H}_{m_{2}}(z) \, , \\
& \varphi_{0}^{(2)}(x,t)=\mathcal{N}_{0}^{(m_{1},m_{2})} \frac{e^{i\left( \frac{\sigma\dot{\sigma}}{4}z^{2}-\frac{\sigma\dot{\gamma}}{2}z \right)}}{\sqrt{\sigma}} \frac{e^{-z^2/2}}{g_{m_{1},m_{2}}(z)}\mathcal{H}_{m_{1}}(z) \, ,
\end{aligned}
\label{eq:REX15}
\end{equation}
for $m_{1}=0,2,\cdots$ and $m_{2}=1,3,\cdots$. The polynomials $P_{n}^{(m_{1},m_{2})}(z)$ are defined as
\begin{multline}
P_{n}^{(m_{1},m_{2})}(z)=(m_{2}-m_{1})\mathcal{H}_{m_{1}}(z)\mathcal{H}_{m_{2}}(z)\mathtt{H}_{n+1}(z) \\
2\left[m_{1}(n+m_{2}+1)\mathcal{H}_{m_{1}-1}(z)\mathcal{H}_{m_{2}}(z)-m_{2}(n+m_{1}+1)\mathcal{H}_{m_{1}}(z)\mathcal{H}_{m_{2}-1} \right]\mathtt{H}_{n}(z) \, .
\end{multline}
The normalization constants $\mathcal{N}_{n+1}^{(m_{1},m_{2})}$, for $n=0,1,\cdots$, are determined from~\eqref{eq:MS7}. In analogy with the one-step case, the constant $\mathcal{N}_{0}^{(m_{1},m_{2})}$ is taken from the stationary counterpart~\cite{Mar13}. We thus obtain
\begin{equation}
\begin{aligned}
& \mathcal{N}_{n+2}^{(m_{1},m_{2})}=\left(\sqrt{\pi}2^{n+2}n!(n+m_{1}+1)(n+m_{2}+1)\right)^{-1/2} \, , \\
& \mathcal{N}_{1}^{(m_{1},m_{2})}=\left( \frac{2^{m_{1}-1}m_{1}!}{\sqrt{\pi}(m_{2}-m_{1})}\right)^{1/2} \, , \quad \mathcal{N}_{0}^{(m_{1},m_{2})}=\left(\frac{2^{m_{2}+1}m_{2}!(m_{2}-m_{1})}{\sqrt{\pi}} \right)^{1/2} \, . 
\end{aligned}
\end{equation}

Before concluding this section, let us consider $m_{1}=4$, $m_{2}=5$ and the rest of parameters arbitrary. Straightforward calculations show that the new potential takes the form
\begin{multline}
V_{2}(x,t)=\Omega^{2}(t)x^{2}+F(t)x-\frac{4}{\sigma^{2}} \\
+\frac{64}{\sigma^{2}}\left(\frac{64 z^{14}+ 320 z^{12}+ 528 z^{10} + 1440 z^{8} + 540 z^{6} - 2700 z^4 - 2025 z^2}{256 z^{16}+2048 z^{14}+7936 z^{12}+15360 z^{10}+15840 z^8+5760 z^6+10800 z^4+2025} \right)
\label{eq:REX18}
\end{multline}
where as usual $z(x,t)=(x+\gamma(t))/\sigma(t)$.

\section{Some applications}
\label{sec:EXA}
The results obtained so far has been developed as general as possible. For completeness, we consider some specific profiles for the frequency term and the external driving force of the initial parametric oscillator~\eqref{eq:PO1}. Time-dependent models are useful in the study of electromagnetic traps of charged particles, from which the parametric oscillator emerges naturally as a suitable model to characterize those systems~\cite{Pau90}. The analysis presented in~\cite{Gla92} reveals that frequency terms which are both even and periodic functions of time lead to localizable probability distributions, that is, wave-packets constrained to move inside a bounded region in space. The latter is the criteria used to characterize the trapping of particles. 

We thus consider a constant frequency term and a sinusoidal driving force as a first example. In this way we construct, under some given conditions, periodic potentials that meet the trapping condition. A second example is provided by a smooth and non-periodic function, together with a null driving force, such that the trapping condition is achieved, even when the frequency is not a periodic function.

In the rest of the text, we consider the one-step potential $V_{1}(x,t)$ with $m=4$ and the two-step potential $V_{2}(x,t)$ with $\{ m_{1}=4, m_{2}=5\}$.

\subsection{$\Omega^{2}(t)=1$ and $F(t)=F_{0}\cos\alpha t$}
\label{subsec:CF}

\begin{figure}[h]
\centering
\subfloat[][]{\includegraphics[width=0.3\textwidth]{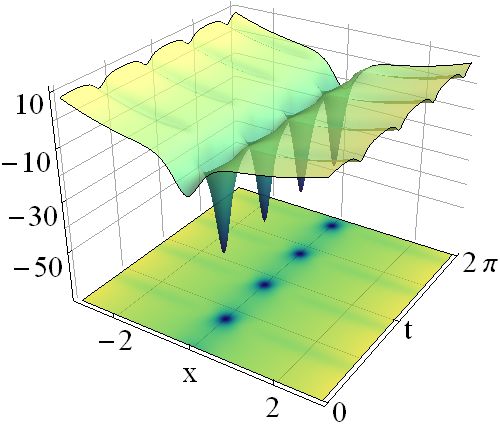}
\label{fig:F1b}}
\hspace{3mm}
\subfloat[][]{\includegraphics[width=0.3\textwidth]{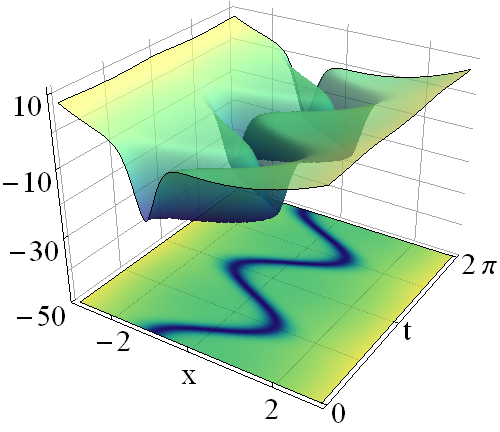}
\label{fig:F1c}}
\hspace{3mm}
\subfloat[][]{\includegraphics[width=0.3\textwidth]{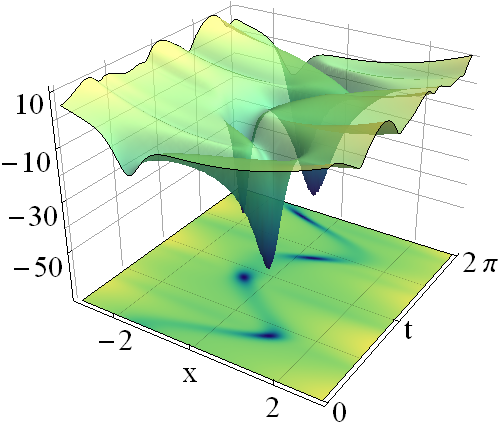}
\label{fig:F1a}}
\\
\subfloat[][]{\includegraphics[width=0.3\textwidth]{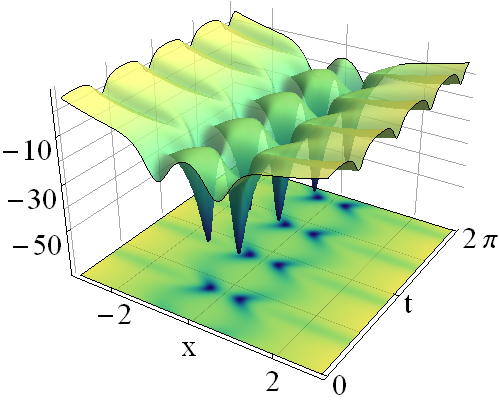}
\label{fig:F1e}}
\hspace{3mm}
\subfloat[][]{\includegraphics[width=0.3\textwidth]{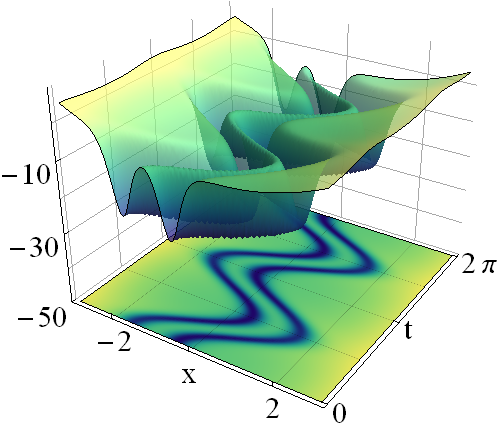}
\label{fig:F1f}}
\hspace{3mm}
\subfloat[][]{\includegraphics[width=0.3\textwidth]{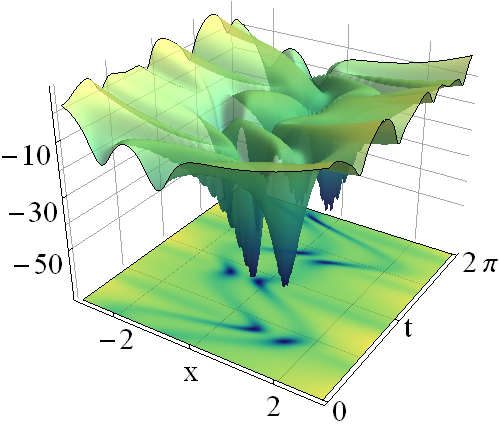}
\label{fig:F1d}}
\caption{Time-dependent potentials associated with the one-step $V_{1}(x,t)$ (upper row) and two-step $V_{2}(x,t)$ (lower row) rational extensions of the parametric oscillator for $m=4$ and $\{m_{1}=4,m_{2}=5\}$, respectively. The set of parameters $\{\mathcal{A},\phi,a,c,F_{0},\alpha \}$ is fixed to $\{0,0,\sqrt{2},\sqrt{2},0,3 \}$ (left column), $\{1,0,1,1,0,3 \}$ (middle column) and $\{1,0,\sqrt{2},\sqrt{2},1,3 \}$ (right column).}
\label{fig:F1}
\end{figure}
In this case, the solutions of the Ermakov equation and $\gamma(t)$ are computed from~\eqref{eq:PO4} with ease, leading to
\begin{equation}
\begin{aligned}
& z(x,t)=\frac{x+\mathcal{A}\cos(2t+\phi)+\gamma_{p}(t)}{\left( \frac{a+c}{2}+\frac{a-c}{2}\cos 4t + \sqrt{ac-1} \, \sin 4t \right)^{1/2}} \, , \quad \mathcal{A},F_{0},\alpha\in\mathbb{R} \, , \quad \phi\in[0,2\pi] \, , \\
&  \gamma_{p}(t)=\begin{cases} \frac{2F_{0}}{4-\alpha^{2}}\cos\alpha t & \alpha\neq 2 \\ \frac{F_{0}}{2}t\sin 2t & \alpha=2 \end{cases}
\end{aligned}
\label{eq:CF2}
\end{equation}
The explicit form of the potentials for the one-step and two-step factorizations is given in~\eqref{eq:REX11} and~\eqref{eq:REX18}, respectively. From~\eqref{eq:CF2}, we can classify the behavior of the new potentials and the respective solutions in several classes. 

\begin{figure}[h]
\centering
\subfloat[][$n=0$]{\includegraphics[width=0.3\textwidth]{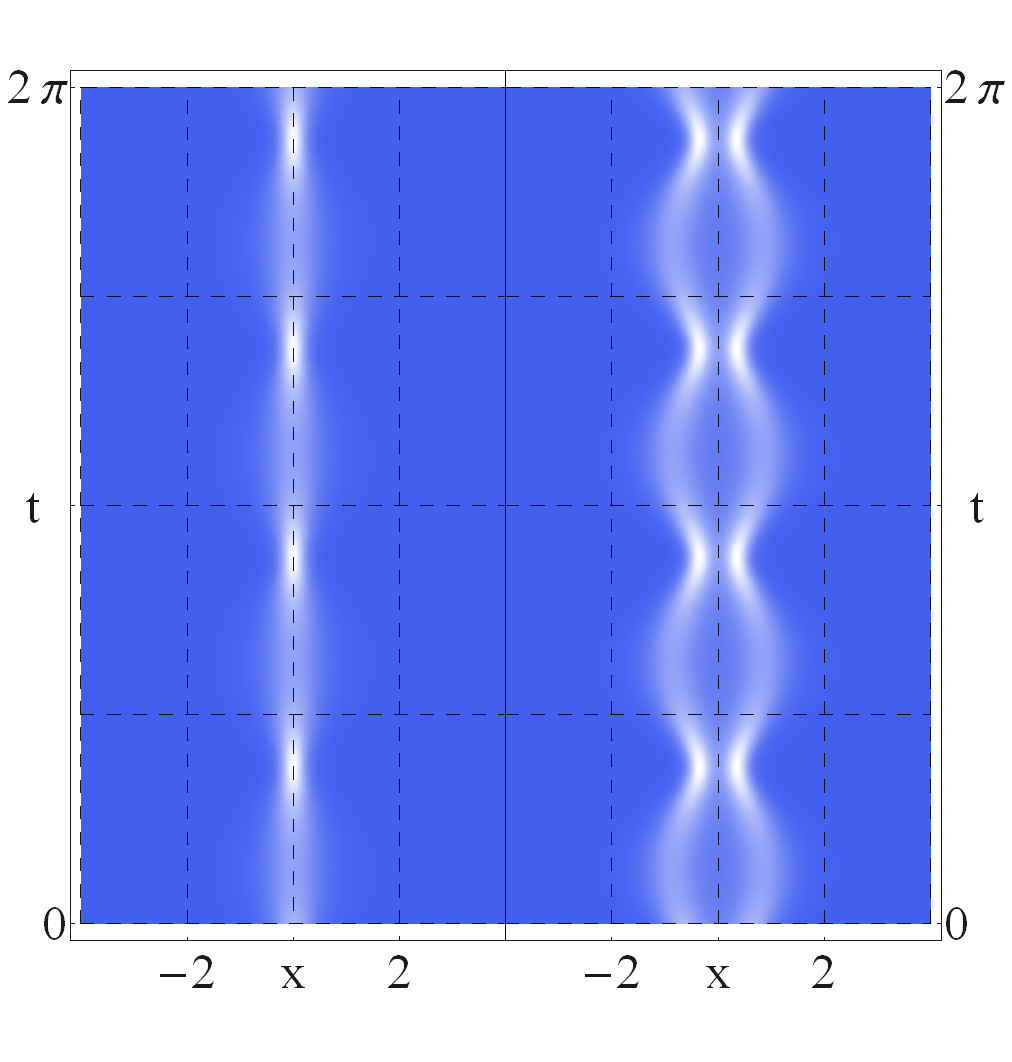}
\label{fig:F2b}}
\hspace{3mm}
\subfloat[][$n=0$]{\includegraphics[width=0.3\textwidth]{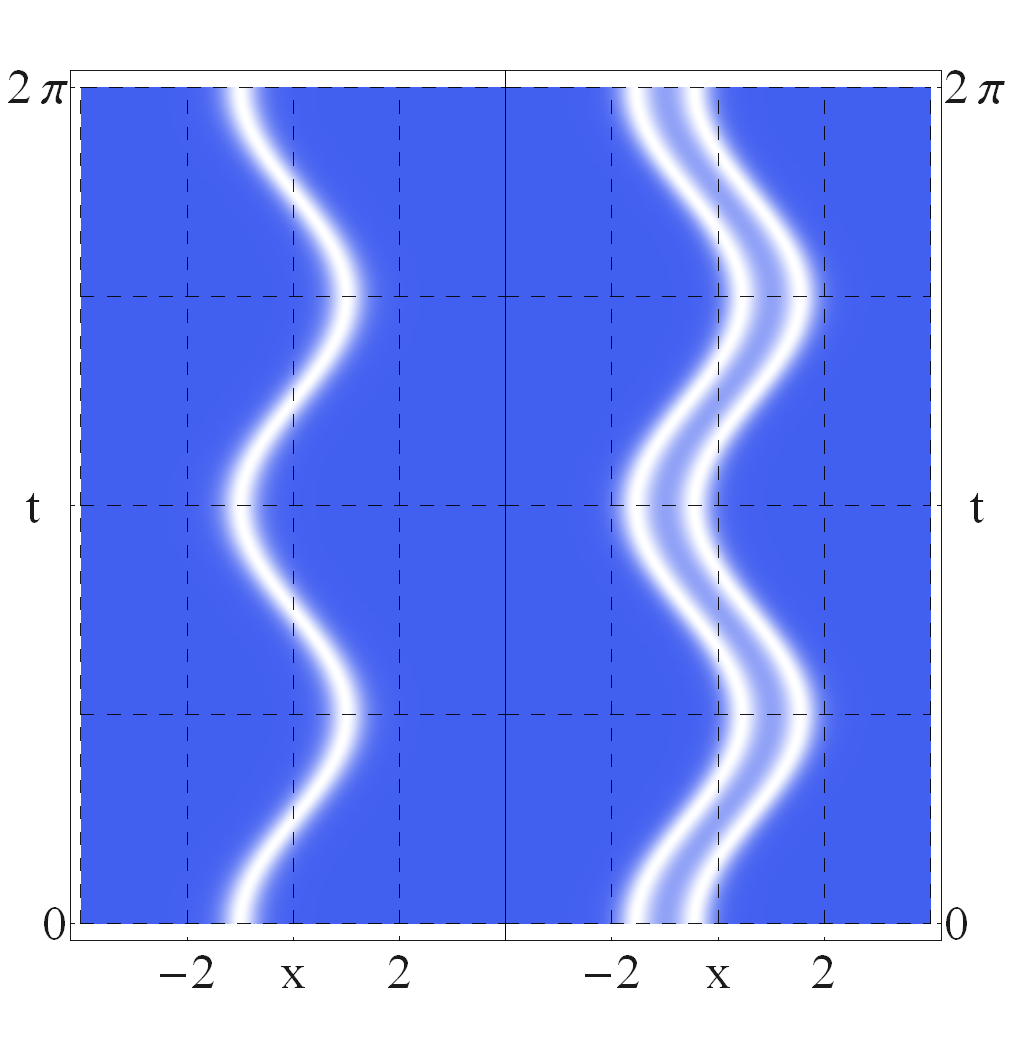}
\label{fig:F2c}}
\hspace{3mm}
\subfloat[][$n=0$]{\includegraphics[width=0.3\textwidth]{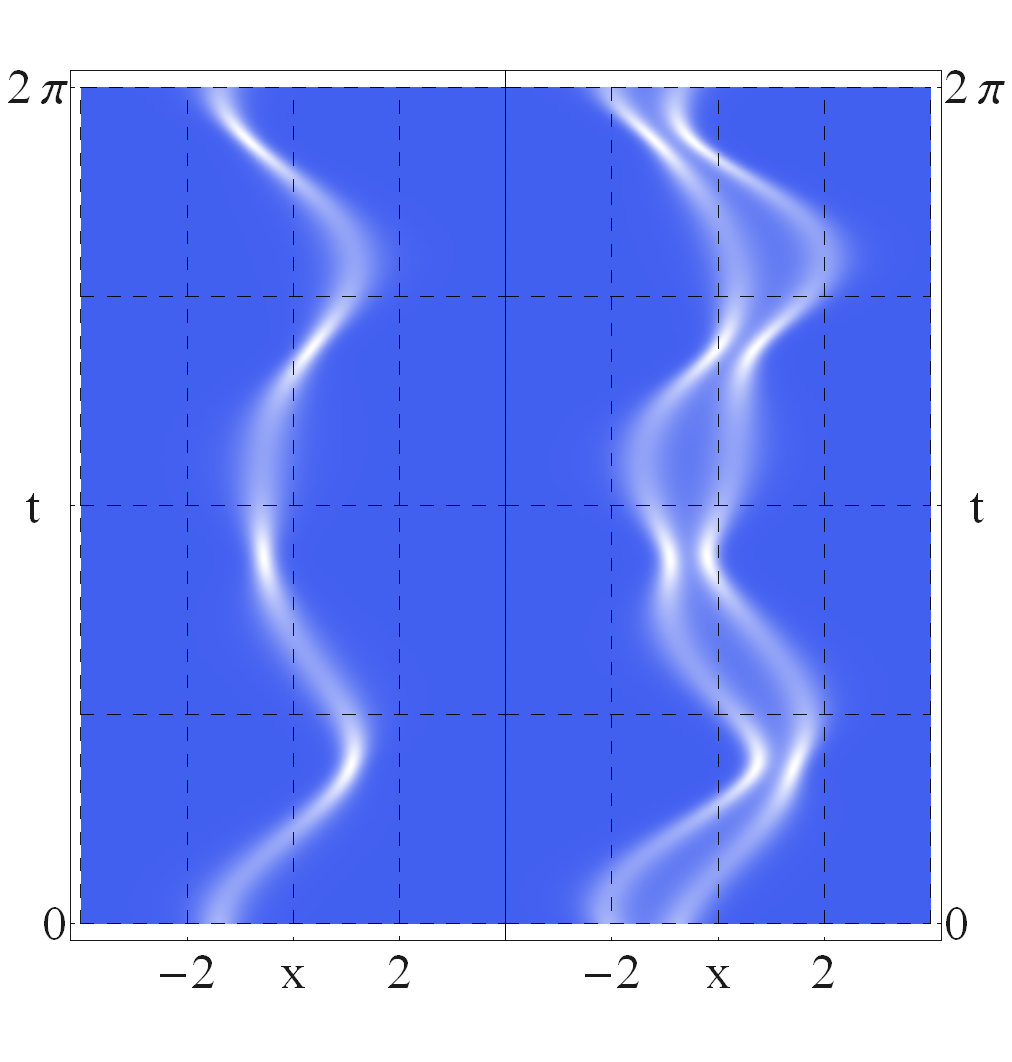}
\label{fig:F2a}}
\\
\subfloat[][$n=1$]{\includegraphics[width=0.3\textwidth]{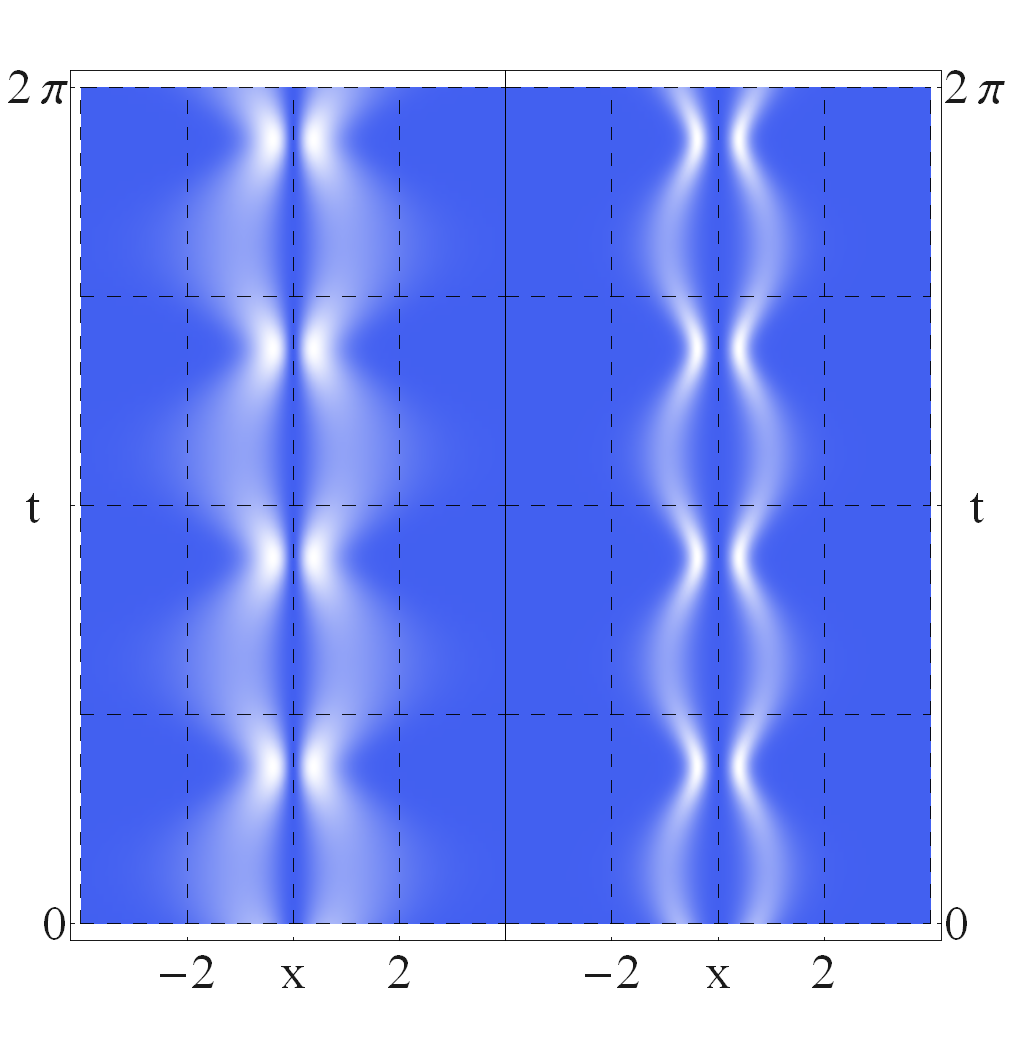}
\label{fig:F2e}}
\hspace{3mm}
\subfloat[][$n=1$]{\includegraphics[width=0.3\textwidth]{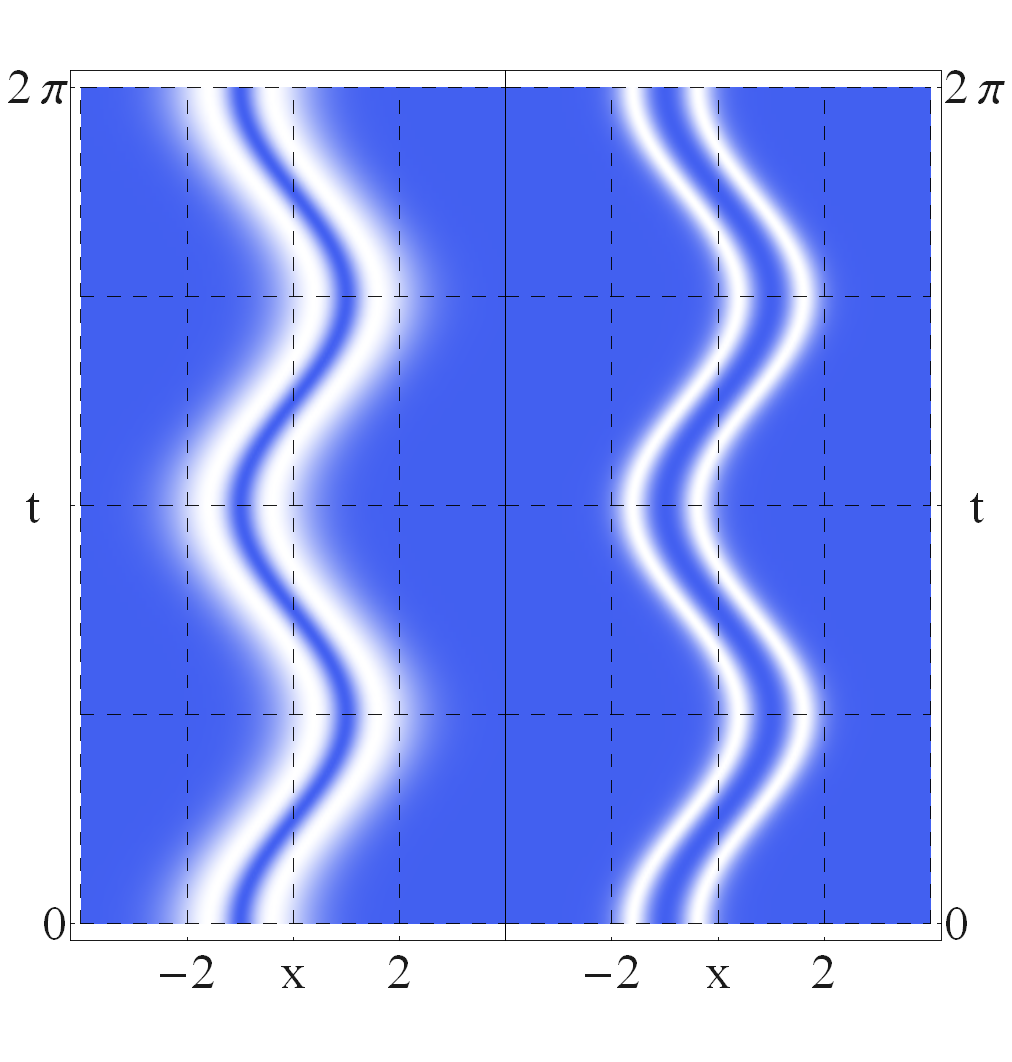}
\label{fig:F2f}}
\hspace{3mm}
\subfloat[][$n=1$]{\includegraphics[width=0.3\textwidth]{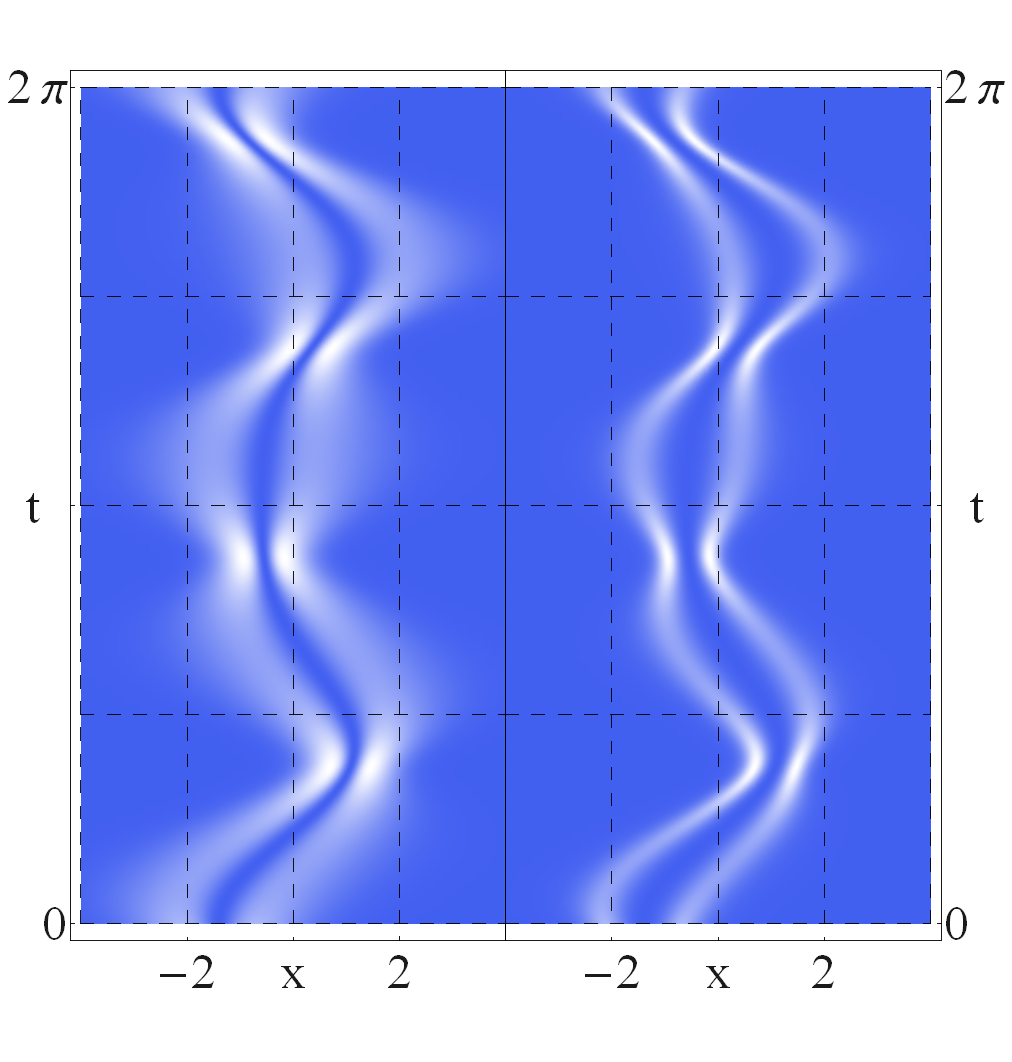}
\label{fig:F2d}}
\caption{Probability distribution $\vert\psi_{n}^{(1)}(x,t)\vert^2=\vert\varphi_{n}^{(1)}(x,t)\vert^2=$ \eqref{eq:REX10} (left-side in each panel) and $\vert\psi_{n}^{(2)}(x,t)\vert^2=\vert\varphi_{n}^{(2)}(x,t)\vert^2=$ \eqref{eq:REX15} (right-side of each panel). For the one-step we have used $m=4$ and for the two-step $\{ m_{1}=4,m_{2}=5 \}$. The set in parameters $\{\mathcal{A},\phi,a,c,F_{0},\alpha \}$ is fixed to $\{0,0,\sqrt{2},\sqrt{2},0,3 \}$ (left column), $\{1,0,1,1,0,3 \}$ (middle column) and $\{1,0,\sqrt{2},\sqrt{2},1,3 \}$ (right column).}
\label{fig:F2}
\end{figure}

For $F_{0}=0$ we recover the harmonic oscillator limit, that is, the initial Hamiltonian $H_{0}(t)$ reduces to the stationary oscillator Hamiltonian $H=$\eqref{eq:HO1}. Interestingly, the resulting potentials $V_{1,2}(x,t)$ are in general time-dependent, even though the initial model is stationary. The time dependence is inherited from the invariant operator $I_{0}(t)$ to the new invariant $I_{1}(t)$ and consequently to the Hamiltonian $H_{1}(t)$. It means that there is a clear difference between our approach and the conventional factorization. Moreover, the potentials $V_{1,2}(x,t)$ are in this case periodic functions of time. The periodicity depends on the values of the parameters $a,c,\mathcal{A}$. If $a,c\neq 1$ and $\mathcal{A}\in\mathbb{R}$, the periodicity becomes $T=\pi/2$ and the initial solutions $\psi^{(0)}_{n}(x,t)$ reduce to the squeezed number states, discussed in Sec.~\ref{subsec:GCS}. To illustrate the latter, we depict the behavior of the new potentials $V_{1,2}(x,t)$ for $\mathcal{A}=0$ in Fig.~\ref{fig:F1b} and Fig.~\ref{fig:F1e}, respectively.  The one-step factorization leads to a deformed oscillator with one minimum localized at $x=0$, where the depth and width of the deformation is changing in time in a periodic way. In turn, the two-step factorization produces a potential with two moving minima. The respective probability densities are depicted in Fig.~\ref{fig:F2b} and Fig.~\ref{fig:F2e}.

If we consider $F_{0}=0$ and $a=c=1$, the periodicity of the new potentials become $T=\pi$. Their behavior is depicted in Fig.~\ref{fig:F1c} for the one-step and in Fig.~\ref{fig:F1d} for the two-step factorization. From these pictures it is clear that the depth and width of the deformations are preserved at each time, nevertheless, the minima position is moving in time. The latter can be seen from the explicit form of the potentials and the fact that $\sigma(t)=1$. Also, it is worth to recall that in this case the initial solutions $\psi_{n}^{(0)}(x,t)$ reduce to the generalized coherent states~\cite{Phi14}. Therefore, we call to the new solutions obtained from the one-step and two-step factorizations as \textit{the rational extensions of the generalized coherent states}. The respective probability densities are depicted in Figs.~\ref{fig:F2c}, \ref{fig:F2f}. 

The conventional stationary rational extensions of the harmonic oscillator are recovered in the special case $a=c=1$ and $F_{0}=\mathcal{A}=0$. In this case we have $\sigma(t)=1$, $\gamma(t)=0$ and $z=x$. In this way the time-dependence is removed from the initial parametric oscillator and from the new rational extensions, leading to the same potentials reported previously in the literature~\cite{Mar13,Gom14}.

Finally, we consider the case in which the driving force acts on the system, $F_{0}\neq 0$. The periodicity of the new potentials $V_{1,2}(x,t)$ can not be taken for granted. By inspecting the reparametrized variable $z(x,t)$ it is clear that periodic potentials are achieved only if $\alpha$ is a rational number and $\alpha\neq 2$, say $\alpha=n_{1}/n_{2}$ with $n_{1},n_{2}\in\mathbb{Z}/\{0\}$. Thus, the periodicity is manipulated by tuning $\alpha$. A particular example is presented in the potentials $V_{1,2}(x,t)$ of Fig.~\ref{fig:F1a} and Fig.~\ref{fig:F1d} for $\alpha=3$, where the periodicity becomes $T=2\pi$. The respective probability densities are shown in Figs.~\ref{fig:F2a}, \ref{fig:F2d}. On the other hand, the resonant case $\alpha=2$ is clearly non-periodic, since $\gamma_{p}(p)=$\eqref{eq:CF2} has a linear term in $t$. Moreover, the dynamics of the minima becomes unbounded as time passes. Such a behavior is not desired if we are looking for localized wave-packets constrained to move in a bounded region, such as the ones depicted in Fig.~\ref{fig:F2}.

\subsection{$\Omega^{2}(t)=\Omega_{1}+\Omega_{2}\operatorname{sech}^{2}k(t-t_{0})$ and $F(t)=0$}
\label{subsec:SF}

\begin{figure}[h]
\centering
\subfloat[][$m=4$]{\includegraphics[width=0.4\textwidth]{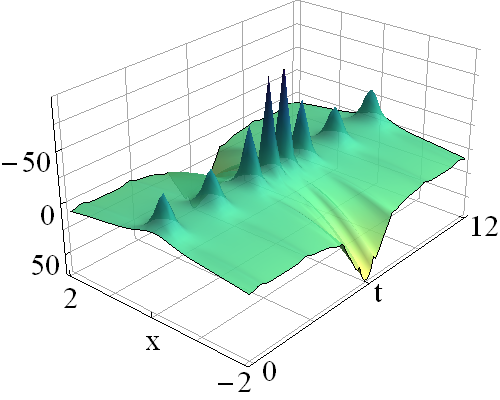}
\label{fig:FSHa}}
\hspace{2mm}
\subfloat[][$m_{1}=4$, $m_{2}=5$]{\includegraphics[width=0.4\textwidth]{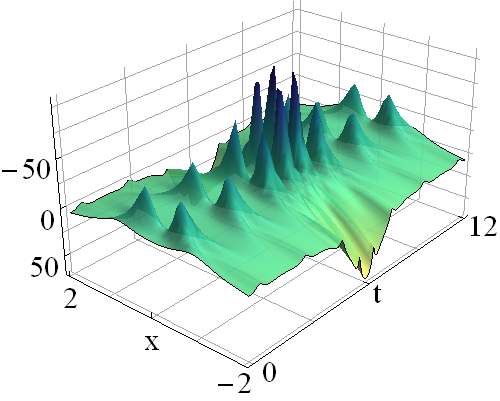}
\label{fig:FSHb}}
\\
\subfloat[][$n=0$]{\includegraphics[width=0.3\textwidth]{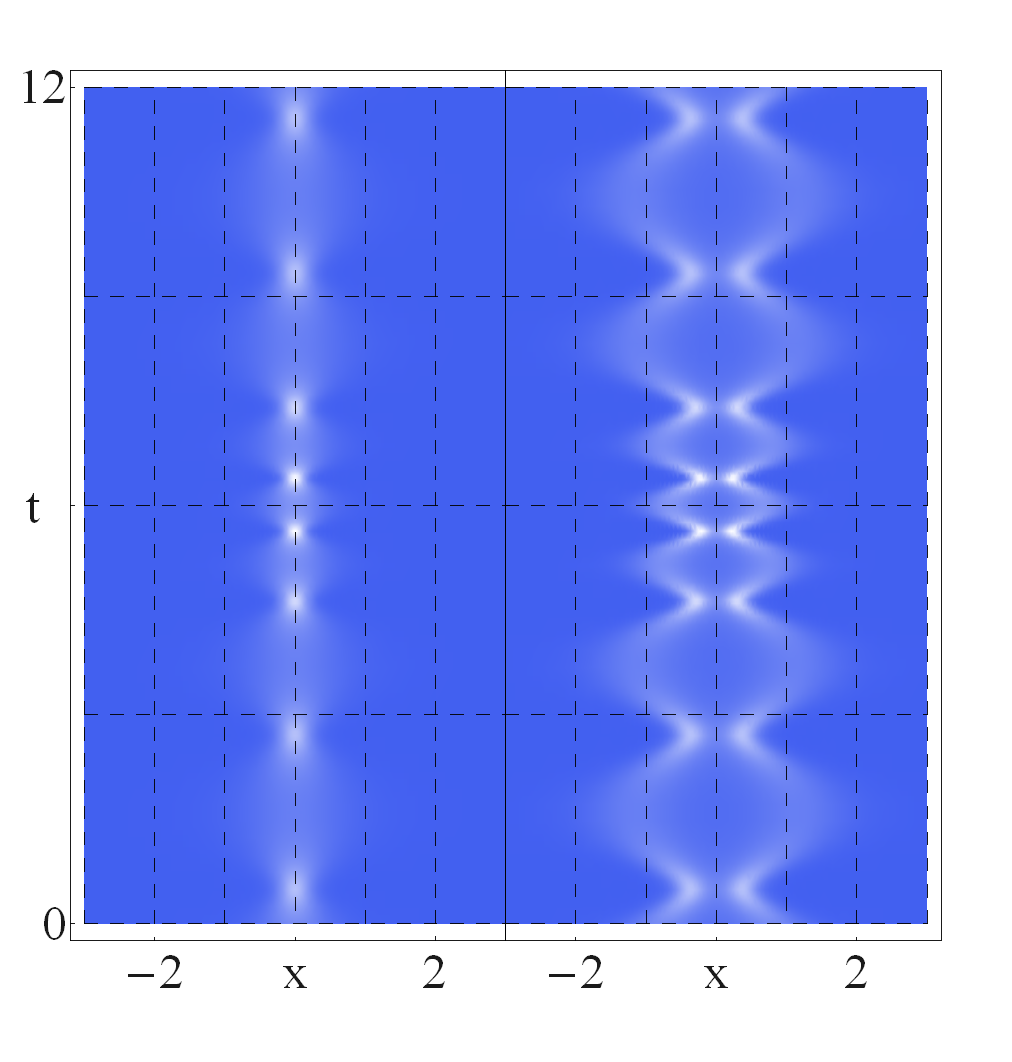}
\label{fig:FSHc}}
\hspace{2mm}
\subfloat[][$n=1$]{\includegraphics[width=0.3\textwidth]{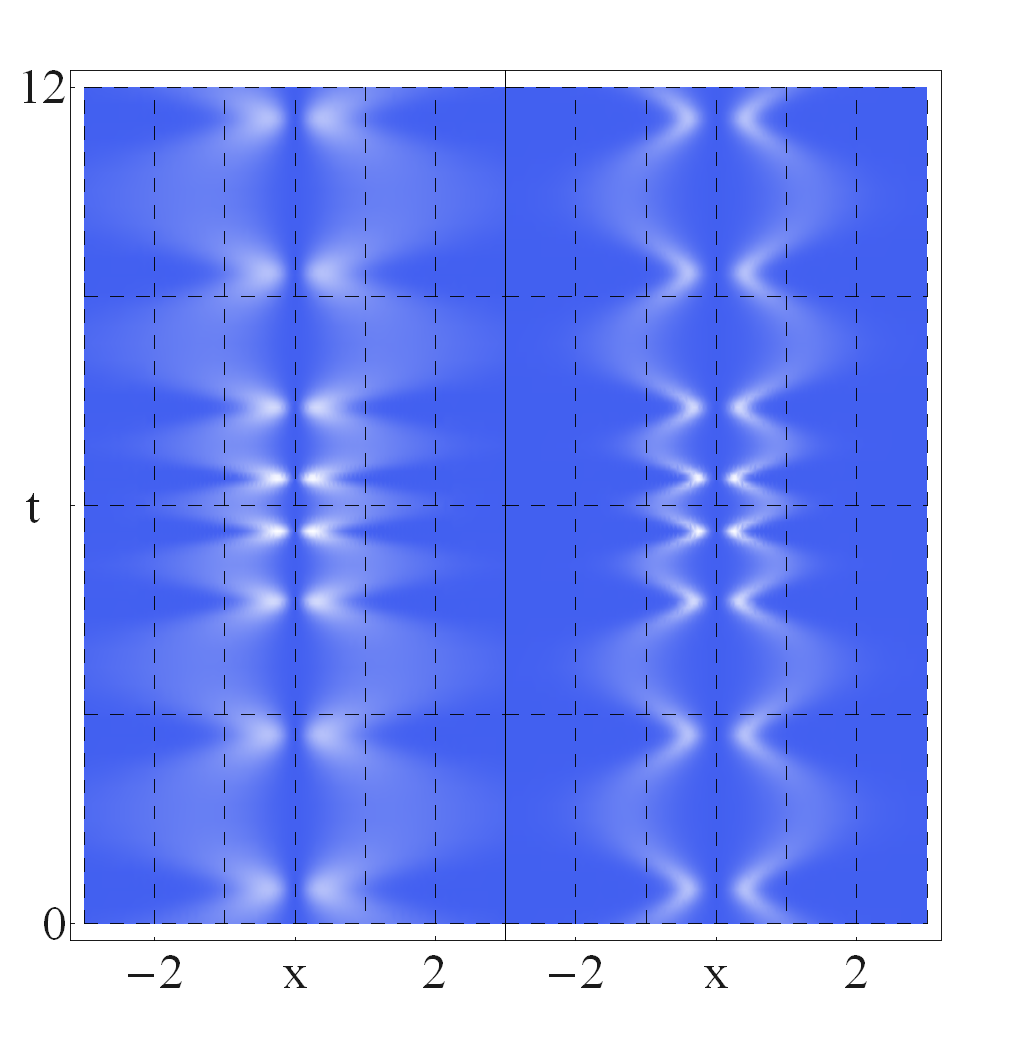}
\label{fig:FSHd}}
\caption{(First row) One-step and two-step rational extensions $V_{1}(x,t)$~(a) and $V_{2}(x,t)$~(b) for the frequency $\Omega^{2}(t)=\Omega_{1}+\Omega_{2}\operatorname{sec}^{2}k(t-t_{0})$, null driving force ($F(t)=0$) and $a=c=k=1, \gamma_{1}=\gamma_{2}=0, \Omega_{1}=2, \Omega_{2}=15, t_{0}=6$. The potentials depicted in panels (a) and (b) are presented upside-down to ease their visualization. (Second row) Probability density $\vert\psi_{n}^{(1)}(x,t)\vert^{2}$ (left-side) and $\vert\psi_{n}^{(2)}(x,t)\vert^{2}$ (right-side) for the potentials depicted in (a) and (b).}
\label{fig:FSH}
\end{figure}

The restriction on the constants $\Omega_{1},\Omega_{2}>0$ is imposed to guarantee $\Omega^{2}(t)>0$ at each time. This frequency profile behaves like a constant $\Omega_{1}$ for $\vert k(t-t_{0})\vert\gg 1$ and changes smoothly to reach its maximum value $\Omega_{1}+\Omega_{2}$ at $t=t_{0}$. In the limit $k\rightarrow 0$, with $\Omega_{2}=1/k$ and $\Omega_{1}=0$, the frequency $\Omega^{2}(t)$ converges to a Dirac-delta distribution $\delta(t-t_{0})$. For simplicity, throughout this section we consider a null driving force. We thus have
\begin{equation}
\begin{aligned}
& q_{1}(t)=(1-y)^{-i\frac{\sqrt{\Omega_1}}{2k}}(1+y)^{i\frac{\sqrt{\Omega_1}}{2k}} \, {}_{2}F_{1}\left( \left. \begin{aligned} \frac{1}{2} + \nu \hspace{2mm} , \hspace{2mm} \frac{1}{2}-\nu \\ 1-i\frac{\sqrt{\Omega_1}}{k} \hspace{5mm} \end{aligned} \right\vert \frac{1-y}{2} \right) \, , \\
& q_{2}(t)=(1-y)^{i\frac{\sqrt{\Omega_1}}{2k}}(1+y)^{-i\frac{\sqrt{\Omega_1}}{2k}} \, {}_{2}F_{1}\left( \left. \begin{aligned} \frac{1}{2} + \nu \hspace{2mm} , \hspace{2mm} \frac{1}{2}-\nu \\ 1+i\frac{\sqrt{\Omega_1}}{k} \hspace{5mm} \end{aligned} \right\vert \frac{1-y}{2} \right) \, ,
\end{aligned}
\label{eq:SF1}
\end{equation} 
with $y=\tanh k(t-t_{0})$, $\nu=\sqrt{\frac{1}{4}+\frac{\Omega_2}{k^2}}$ and the Wronskian $W(q_{1},q_{2})=w_{0}=-2i\sqrt{\Omega_1}$. Notice that $q_{1,2}$ are both complex-valued functions, where $q_{2}^{*}=q_{1}$. To obtain a real-valued function $\sigma(t)$ we set $a=c$ in~\eqref{eq:PO41}, leading to
\begin{equation}
\begin{aligned}
\sigma^{2}(t)&=2a\operatorname{Re}[q_{1}^{2}(t)]+2\sqrt{a^2+\frac{1}{\Omega_1}} \, \vert q_{1}(t)\vert^{2} \\
&=2\left(a+\sqrt{a^{2}+\frac{1}{\Omega_{1}}} \right)\operatorname{Re}[q_{1}(t)]^{2}+2\left( -a+\sqrt{a^{2}+\frac{1}{\Omega_{1}}} \right)\operatorname{Im}[q_{1}(t)]^{2} \, .
\end{aligned}
\label{eq:SF2}
\end{equation}
From the latter result it is clear that $\sigma(t)$ is a nodeless function for $a\neq 0$, as required to obtain a regular potential and solutions at each time. Given the absence of the driving force, the nonhomogeneous equation for $\gamma$ reduces to a homogeneous one, which has as a solution
\begin{equation}
\gamma(t)=\gamma_{1}\operatorname{Re}[q_{1}(t)]+\gamma_{2}\operatorname{Im}[q_{1}(t)] \, ,
\label{eq:SF3}
\end{equation}
with $\gamma_{1}$ and $\gamma_{2}$ arbitrary real constants to guarantee that $\gamma(t)$ is a real-valued function.

Following~\eqref{eq:SF1}, it is clear that the potentials $V_{1,2}(x,t)$ are in general non-periodic functions of time. Although, for asymptotic times $\vert k(t-t_{0})\vert \gg 1$ the potentials can be approximated to periodic functions. To illustrate our results, we depict $V_{1}(x,t)$ in Fig.~\ref{fig:FSHa} and $V_{2}(x,t)$ in Fig.~\ref{fig:FSHb}. In the one-step case, the potential has a global minimum that oscillates in time in a non-periodic way, those oscillations increase and reach their maximum around $t\approx t_{0}$, then, for asymptotic time the $\vert k(t-t_{0})\vert\gg 1$ the wave-packets approximate to a periodic function. Equivalent information is extracted from the probability densities of Figs.~\ref{fig:FSHc}-\ref{fig:FSHd}, from where highly localizable wave-packets are observed at times $t\approx t_{0}$.

\section{Conclusions}
We have shown the proper way to implement the conventional factorization method to the quantum invariant of the parametric oscillator. This led to a new family of quantum invariants and their respective time-dependent Hamiltonians. The spectral properties of the new invariant are inherited from the initial one, and thus the solutions to the Schr\"odinger equation are determined by applying the proper mappings. In the constant frequency case, in general, the quantum invariant preserve the time dependence and it becomes a constant of motion of the stationary oscillator. Moreover, the eigenfunctions of such an invariant reduce to either the squeezed or displaced number states~\cite{Nie97}. Thus, those states are nothing but the eigenfunctions of the appropriate quantum invariant of the stationary oscillator. This is a connection that, to the authors knowledge, has not been noticed previously. From the latter, the rational extensions of the displaced or squeezed number states follow as a special case. Thus, we obtain more generality by considering the factorization method with the quantum invariants, for it leads to new time-dependent Hamiltonians, even in the harmonic oscillator limit. Nevertheless, the conventional stationary factorization is recovered in  a very special limit.

This construction can be extended to other exactly solvable time-dependent models such as the singular oscillator~\cite{Dod98} and the Caldirola-Kanai oscillator~\cite{Cal41,Kan48,Gue15}. Moreover, it is worth exploring the non-Hermitian models in the $\mathcal{PT}$ regime such as the Swanson oscillator~\cite{Swa04,Fri16,Bag15}, as well as the non-$\mathcal{PT}$ models constructed and studied in~\cite{Ros15,Ros18}. The latter deserve a special treatment and will be discussed in detail elsewhere.

\section*{Acknowledgments}
K. Zelaya is supported by the Mathematical Physics Laboratory, Centre de Recherches Math\'ematiques, through a postdoctoral fellowship. He also acknowledges the support of Consejo Nacional de Ciencia y Tecnolog\'ia (Mexico), grant number A1-S-24569. V. Hussin acknowledges the support of research grants from NSERC of Canada.

\appendix

\setcounter{section}{0}  
\section{Computing  $\chi^{(k)}_{n}(t)$}
\label{sec:APPC}
\renewcommand{\thesection}{A-\arabic{section}}

\renewcommand{\theequation}{A-\arabic{equation}}
\setcounter{equation}{0}  
The complex-phase $\chi^{(0)}_{n}(t)$ that connects the eigenfunction of the invariant operator $\varphi^{(0)}_{n}(x,t)$ into solution of the Schr\"odinger equation is computed from
\begin{equation}
\dot{\chi}^{(0)}_{n}=\langle\varphi_{n}^{(0)}(t)\vert i\frac{\partial}{\partial t}-H_{0}(t)\vert\varphi_{n}^{(0)}(t)\rangle \, .
\label{eq:APPC0}
\end{equation}
The calculation of~\eqref{eq:APPC0} is not trivial, since $\varphi_{n}^{(0)}(x,t)$ is not an eigenfunction of $H_{0}(t)$. Nevertheless, we can use the reparametrization $z(x,t)$ introduced in~\eqref{eq:SPO4} to rewrite~\eqref{eq:APPC0} in terms of the invariant $I_{0}(t)$, for which $\varphi_{n}^{(0)}(x,t)$ is an eigenfunction. With the use of the reparametrization $z(x,t)$, the eigenvalue equation associated with the invariant operator $I_{0}$ takes the form
\begin{multline}
I_{0}(t)\varphi_{n}^{(0)}=-\frac{\partial^2\varphi_{n}^{(0)}}{\partial z^{2}}+\left(\frac{\sigma^{2}\dot{\sigma}^{2}}{4}+1 \right)z^{2}\varphi_{n}^{(0)}-\frac{\sigma\dot{\sigma}}{2i}\left[2\left(z-\frac{\gamma}{\sigma} \right)\frac{\partial}{\partial z}+1 \right]\varphi_{n}^{(0)} \\
+\frac{\sigma\dot{\gamma}-\dot{\sigma}\gamma}{i}\frac{\partial\varphi_{n}^{(0)}}{\partial z}-\frac{\sigma^{2}\dot{\sigma}\dot{\gamma}}{2}z\varphi_{n}^{(0)}+\frac{\dot{\gamma}^{2}\sigma^{2}}{4}\varphi_{n}^{(0)}=(2n+1)\varphi_{n}^{(0)} \, .
\label{eq:APPC1}
\end{multline}
In addition, the use of the chain differentiation rule allows us to rewrite the action of the partial derivative with respect to time as
\begin{multline}
i\frac{\partial\varphi_{n}^{(0)}}{\partial t}=\left[ -\left(\frac{\ddot{\sigma}\sigma+\sigma^{2}}{4}\right)z^{2}+\left(\frac{\sigma\ddot{\gamma}+\dot{\gamma}\dot{\sigma}}{2}\right)z-\left(\frac{\gamma\ddot{\gamma}+\dot{\gamma}^{2}}{2}\right) \right]\varphi_{n}^{(0)} \\
+\frac{\dot{\sigma}}{2i\sigma}\left[2\left(z-\frac{\gamma}{\sigma} \right)\frac{\partial}{\partial z}+1 \right]\varphi_{n}^{(0)}-\frac{\sigma\dot{\gamma}-\dot{\sigma}\gamma}{i\sigma^{2}}\frac{\partial\varphi_{n}^{(0)}}{\partial z} \, .
\label{eq:APPC2}
\end{multline}
The action of the Hamiltonian $H_{0}$ takes the form
\begin{equation}
H_{0}(t)\varphi_{n}^{(0)}=-\frac{1}{\sigma^{2}}\frac{\partial^{2}\varphi_{n}^{(0)}}{\partial z^{2}}+\left[\Omega^{2}\sigma^{2}z^{2}-\left(2\sigma\gamma\Omega+F\sigma \right)z + \left( \Omega^{2}\gamma^{2}-F\gamma \right)\right]\varphi_{n}^{(0)} \, . 
\label{eq:APPC3}
\end{equation}
By combining Eqs.~\eqref{eq:APPC2}-\eqref{eq:APPC3} and using the differential equations for $\sigma$ and $\gamma$ in~\eqref{eq:PO4} we end up with the following simple expression:
\begin{equation}
\left(i\frac{\partial}{\partial t}-H_{0}(t)\right)\varphi_{n}^{(0)}=\left(-\frac{1}{\sigma^{2}}I_{0}(t)-\frac{\dot{\gamma}^{2}+\gamma\ddot{\gamma}}{4}+\frac{\gamma F}{2} \right)\varphi_{n}^{(0)} \, .
\label{eq:APPC4}
\end{equation}
From~\eqref{eq:APPC4} and from the fact that the eigenfunctions $\varphi_{n}^{(0)}(x,t)$ are already normalized and the physical inner product~\eqref{eq:PO72} does not depend on time,  we obtain the final result
\begin{equation}
\chi^{(0)}_{n}=-(2n+1)\int^{t}\frac{dt'}{\sigma^{2}(t')}-\frac{\gamma\dot{\gamma}}{4}+\frac{1}{2}\int^{t}dt' \, F(t')\gamma(t') \, .
\label{eq:APPC5}
\end{equation}

$\bullet$ \textbf{Cases $\chi_{n}^{(1)}(t)$ and $\chi_{n}^{(k)}(t)$}.
 
From the one-step factorization of Sec.~\ref{subsec:1SF}, the respective complex-phase is computed from
\begin{equation}
\dot{\chi}^{(1)}_{n}=\langle\varphi_{n}^{(1)}(t)\vert i\frac{\partial}{\partial t}-H_{1}\vert\varphi_{n}^{(1)}(t)\rangle \, .
\label{eq:APPC10}
\end{equation}
With the use of the mapped eigenfunctions\footnote{The same result holds for the missing state $\varphi_{\epsilon_{1}}^{(1)}$.} $\varphi_{n+1}^{(1)}\propto B_{1}\varphi_{n}^{(0)}$ given in~\eqref{eq:SPO12} and the reparametrization $z(x,t)$ we obtain 
\begin{equation}
\left(i\frac{\partial}{\partial t}-H_{1}(t)\right)\varphi_{n+1}^{(1)}=\left(-\frac{1}{\sigma^{2}}I_{1}(t)-\frac{1}{4}\frac{d}{dt}\gamma\dot{\gamma}+\frac{\gamma F}{2} \right)\varphi_{n+1}^{(1)} \, ,
\label{eq:APPC12}
\end{equation}
from which we extract the complex-phase as
\begin{equation}
\chi^{(1)}_{n}=-\lambda_{n}^{(1)}\int^{t}\frac{dt'}{\sigma^{2}(t')}-\frac{\gamma\dot{\gamma}}{4}+\frac{1}{2}\int^{t}dt'\gamma(t')F(t') \, .
\label{eq:APPAC13}
\end{equation}
For factorization of higher order the procedure is quite similar, leading to
\begin{equation}
\chi^{(k)}_{n}=-\lambda_{n}^{(k)}\int^{t}\frac{dt'}{\sigma^{2}(t')}-\frac{\gamma\dot{\gamma}}{4}+\frac{1}{2}\int^{t}dt'\gamma(t')F(t') \, ,
\label{eq:APPAC14}
\end{equation}
where $\lambda_{n}^{(k)}$ are the eigenvalues of the invariant operator $I_{k}(t)$.

\appendix
\setcounter{section}{1}  
\section{Determining the time-dependent Hamiltonians}
\label{sec:APPB}
\renewcommand{\thesection}{B-\arabic{section}}

\renewcommand{\theequation}{B-\arabic{equation}}
\setcounter{equation}{0}  

In this section, we show the explicit calculations necessary to determine the time-dependent Hamiltonians generated from the factorization method. Let us consider the one-step case and the ansatz 
\begin{equation}
H_{1}(t)=H_{0}(t)+G(t)\mathfrak{F}(z) \, ,
\end{equation} 
where $G(t)$ and $\mathfrak{F}(z)$ are determined from the quantum invariant condition
\begin{equation}
\frac{d}{dt}I_{1}(t)=i[H_{1}(t),I_{1}(t)]+\frac{\partial}{\partial t}I_{1}(t)=0 \, ,
\label{eq:APPB-1}
\end{equation}
with $I_{1}(t)$ the quantum invariant obtained from the factorization~\eqref{eq:SPO7}. After some calculations, and using the quantum invariant condition for $I_{0}(t)$ given in \eqref{eq:INVI0}, we get the following relationship:
\begin{equation}
iG(t)[\mathfrak{F}(z),I_{0}(t)]+i\left[ H_{0}(t),2\frac{\partial W_{1}}{\partial z} \right]+2\left(\frac{\partial z}{\partial t}\right)\frac{\partial^{2} W_{1}}{\partial z^{2}} \, .
\end{equation} 
The explicit expressions for $H_{0}(t)$ and $I_{0}(t)$ given in~\eqref{eq:PO1} and~\eqref{eq:PO3}, respectively, together with the identity 
\begin{equation}
{}[\hat{x}\hat{p}+\hat{p}\hat{x},f(\hat{x})]=\left. -2ix\frac{\partial f}{\partial x}\right\vert_{x\rightarrow \hat{x}} \, ,
\end{equation} 
for a smooth function $f(x)$, lead us to
\begin{equation}
i\left[\hat{p}^{2},-\sigma^{2}G_{1}(t)\mathfrak{F}_{1}(z)+2\frac{\partial W_{1}}{\partial z} \right]-G_{1}(t)\left(-\dot{\sigma}x+\mathfrak{W} \right)\frac{\partial\mathfrak{F}_{1}}{\partial z}+\frac{2}{\sigma^{2}}\left( -\dot{\sigma}x+\mathfrak{W} \right)\frac{\partial^{2}W_{1}}{\partial z^{2}} =0 \, .
\label{eq:EQU}
\end{equation}
Now, it is clear that $G_{1}(t)=\sigma^{-2}(t)$ and $\mathfrak{F}_{1}(z)=2\partial W_{1}/\partial z$ fulfills~\eqref{eq:EQU}. Therefore, $I_{1}(t)$ is indeed the quantum invariant of the time-dependent Hamiltonian 
\begin{equation}
H_{1}(t)=H_{0}+\frac{2}{\sigma^{2}}\frac{\partial W_{1}}{\partial z} \, .
\end{equation}

For the two-step case, the previous procedure can be extended if we consider the ansatz $H_{2}(t)=H_{1}(t)+G_{2}(t)\mathfrak{F}_{2}(z)$, this leads to $G_{2}(t)=\sigma^{-2}(t)$ and $\mathfrak{F}_{2}(z)=2\partial W_{2}/\partial z$. In the same way, the latter procedure can be generalized for any higher iteration of the factorization method.


\end{document}